\documentclass[]{article}

\usepackage{amsfonts,amsbsy,amscd,amsgen,amsthm,dsfont,amsmath,amssymb,mathrsfs}
\usepackage[colorlinks=true,allcolors=blue]{hyperref}
\usepackage{lipsum}
\usepackage{amsfonts}
\usepackage{graphicx}
\usepackage{epstopdf}
\usepackage[linesnumbered,ruled,vlined]{algorithm2e}
\usepackage{authblk}

\usepackage{placeins}
\usepackage{bbm}

\usepackage{xcolor}

\usepackage{amsopn}

\usepackage{subfigure}
\usepackage{bm} 
\theoremstyle{definition}
\usepackage{enumitem}
\usepackage[utf8]{inputenc}
\usepackage[T1]{fontenc}

\newcommand{\id}{\mathrm d}
\newcommand{\vc}{\mathbf}

\DeclareMathAlphabet\mathbfcal{OMS}{cmsy}{b}{n}

\newcommand{\var}{\mbox{Var}}

\SetKwInput{KwInput}{Input}                
\SetKwInput{KwOutput}{Output} 

\title{Quantifying rare events in spotting: How far do wildfires spread?}
\author{Alex Mendez}
\author{Mohammad Farazmand\thanks{Corresponding author's email address: \href{mailto:farazmand@ncsu.edu}{farazmand@ncsu.edu}}}
\affil{Department of Mathematics, North Carolina State University, 2311 Stinson Drive, Raleigh, NC 27695-8205, USA}
\date{}

\begin{document}
	
	\maketitle

\begin{abstract}
	Spotting refers to the transport of burning pieces of firebrand by wind which, at the time of landing, may ignite new fires beyond the direct ignition zone of the main fire.
	Spot fires that occur far from the original burn unit are rare but have consequential ramifications since their prediction and control remains challenging.
	To facilitate their prediction, we examine three methods for quantifying the landing distribution of firebrands: crude Monte Carlo simulations, importance sampling, and large deviation theory (LDT). 
	In particular, we propose an LDT method that accurately and parsimoniously quantifies the low probability events at the tail of the landing distribution. In contrast, Monte Carlo and importance sampling methods are most efficient in quantifying the high probability landing distances near the mode of the distribution. However, they become computationally intractable for quantifying the tail of the distribution due to the large sample size required. We also show that the most probable landing distance grows linearly with the mean characteristic velocity of the wind field.
	Furthermore, defining the relative landed mass as the proportion of mass landed at a given distance from the main fire, we derive an explicit formula which allows computing this quantity as a function of the landing distribution at a negligible computational cost.
	We numerically demonstrate our findings on two prescribed wind fields.
\end{abstract}

\section{Introduction}
As the amount of human dwellings near forest fire danger zones increases and as climate change results in more amenable conditions to the creation of fires, there is a greater urgency to predict wildfire dynamics and in particular spot fires~\cite{Tedim2018,tohidi2017,Mendez2021,burke2021}.
Forest fires generate burning pieces of vegetation, called firebrands, and launch them into the air through columns of gas produced by the fire. After lofting into the air, firebrands are taken away by the ambient wind. Once landed, these burning firebrands can start separate fires away from the original fire. The secondary fires are called spot fires and the entire process is referred to as spotting~\cite{albini1979,koo2010,tarifa1965}. 

\begin{figure}
	\centering
	\includegraphics[width=0.5\textwidth] {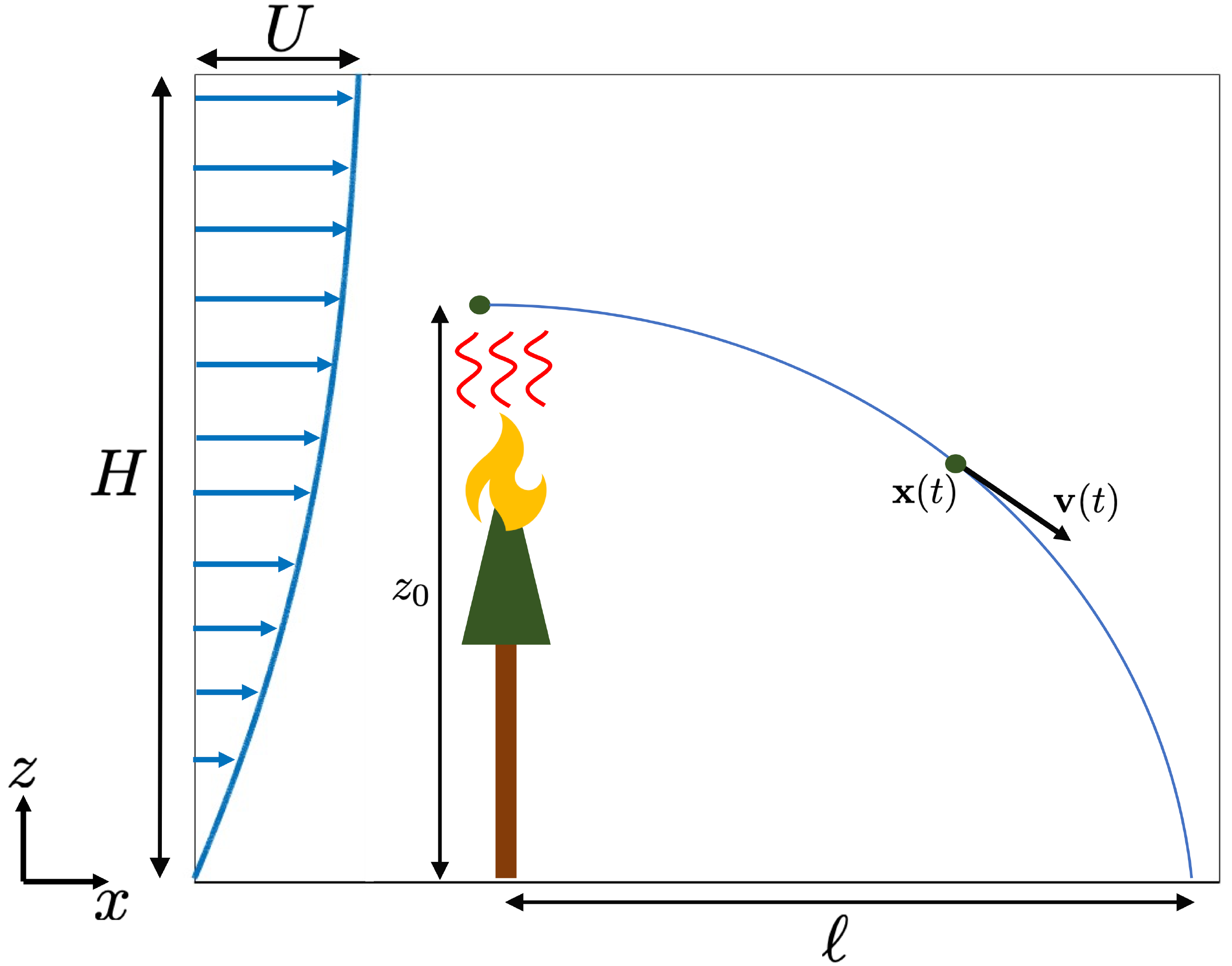}
	\caption{Schematic of firebrand transport from a crown fire. The wind field has asymptotic wind velocity $U$, which is roughly reached at boundary layer height $H$. The firebrand is carried up by the fire's plume to initial height $z_0$, where it then travels due to the wind with position $\vc x(t)$ and velocity $\vc v(t)$. The firebrand eventually lands at landing location $\ell$. }
	\label{fig:schematic}
\end{figure}

The process of spotting consists of three broad phases~\cite{koo2010}. The first is the generation of the firebrands by the main fire. The second is the transport of the firebrands by wind. The third stage occurs after the firebrands land on the ground hence possibly igniting fuels, such as shrubs, dead leaves, and branches, at landing sites (see figure~\ref{fig:schematic}).

In this paper, we focus on the transport stage of the spotting process, particularly when the wind carries the firebrand to its landing location. In particular, given any firebrand-producing fire, we quantify the probability that a firebrand will land at a certain distance from the primary burn unit. We are particularly interested in accurately approximating the probability that a firebrand will land at a distance far away from the original fire, i.e., the tail of the spotting distribution. These are rare but consequential events since spot fires that start far away from the main fire can damage ostensibly safe areas and catch emergency personnel by surprise~\cite{weir2007,koo2010}.

Firebrand transport follows complex nonlinear dynamics arising from the coupling between the firebrand trajectory and the atmospheric flow. Computational fluid dynamics (CFD) models generate the atmospheric flow that transports the firebrands. The firebrand motion is then determined by solving the ordinary differential equations (ODEs) which govern the motion of inertial particles. Monte Carlo simulations of this coupled model can in principle be used to quantify the spotting distribution. However, it is well-known, and we show here in the context of spotting, that Monte Carlo methods with a moderate sample size return only a crude approximation of rare low probability events corresponding to the tail of the distribution~\cite{bucklew2004,farazmand2019a}. In order to obtain a reasonable approximation of the tail an extremely large sample size is required. Given that CFD models are computationally expensive, such Monte Carlo simulations are not practical.

Here, we explore two alternatives: Importance sampling and large deviation theory (LDT). Importance sampling biases the sampling distribution in favor of rare events, hence quantifying the tail of the distribution more accurately while using fewer samples as compared to Monte Carlo methods~\cite{bucklew2004}. 
In contrast, LDT does not require sampling at all; instead, it approximates the tail distribution by an asymptotic expansion determined based on an appropriately defined rate function. 
To evaluate the rate function, one needs to solve an optimization problem which constitutes the main computational cost of LDT. Nonetheless, since LDT does not require sampling, its computational cost is much lower than Monte Carlo and importance sampling methods. Using two prescribed wind fields, we carry out an extensive study, investigating the strengths and drawbacks of Monte Carlo simulation, importance sampling and LDT. While we use classic Monte Carlo and importance methods, the appropriate LDT method was only recently developed~\cite{Dematteis2018,Dematteis2019,tong2021} with its formulation and application to spotting presented here for the first time.

\subsection{Related work}

Much of wildfire research has focused on the physics of coupling between the atmosphere and the fire which involves  fluid dynamics, heat transfer and combustion~\cite{sardoy2007,bhutia2010,jenkins2009,anand2018,manzello2020}.
The resulting CFD models have culminated in comprehensive software packages such as HIGRAD/FIRETEC~\cite{Koo2012}, QUIC-FIRE~\cite{quicfire}, and WRF-SFIRE~\cite{wrfsfire}. 

The study of firebrand trajectories can be traced back to the work of Tarifa et al.~\cite{tarifa1965,Tarifa1963} on the maximum spotting distance of a single firebrand. The maximum spotting distance is the measurement of how far a firebrand can travel between its initial lofting to when it fully burns out.
Using a force balance argument, Tarifa et al.~\cite{tarifa1965,Tarifa1963} used two-dimensional equations of firebrand transport for disk-shaped and cylindrical firebrands. 

In particular, Tarifa et al.~\cite{Tarifa1963} observed that, in a steady and laminar wind field, firebrands reach their terminal velocity after a very short period of time. Consequently, the authors derived an approximation of maximum spotting distance by assuming that a firebrand always moves with its terminal velocity. This assumption subsequently became a common approximation in firebrand research. 

Ever since the pioneering work of Tarifa et al.~\cite{tarifa1965,Tarifa1963}, many similar studies have been conducted which improve or build upon it~\cite{Lee1969,tse1998,zheng2020}. In particular, Koo et al.~\cite{Koo2012} cast doubt on the terminal velocity assumption.
They point out that, although this simplifying assumption is reasonably valid for laminar and steady wind fields, it fails to hold true for more realistic unsteady and inhomogeneous turbulent flows. More specifically, Koo et al.~\cite{Koo2012} find that, in turbulent flows, firebrands without the terminal velocity assumption travel significantly farther than those with this simplifying approximation.

To avoid the computationally expensive firebrand evolution, alternative methods have been proposed~\cite{Albini2012}. For instance, 
cellular automaton (CA) models~\cite{clarke1994,Karafyllidis1997,boychuk2007,Alexandridis2008} seek to estimate the fire propagation on  a specially discrete grid. CA models specify local rules for evolving the fire in discrete time. 
A similar but continuum approach was proposed by Hillen et al.~\cite{Hillen2015} who developed a non-local
transport equation for spotting. Their model expresses the likelihood of fire at a particular time and location in terms of a partial integrodifferential equation (also see Ref.~\cite{Hillen2016}).

Both CA and non-local transport models make several simplifying assumptions, most notably about the wind field and the firebrand trajectories.  Here instead, we estimate the landing distribution of firebrands by directly solving for their trajectories in a given wind field.
The main objective of the present work is to investigate efficient statistical quantification methods for spotting distribution which can ultimately be integrated into existing high-fidelity software packages, such as HIGRAD/FIRETEC, QUIC-FIRE, and WRF-SFIRE. Our main focus is on the tail of the distribution whose statistical quantification presents a major challenge. 
We expect that our findings, specially the accurate quantification of the tail, will also inform and improve existing cellular automaton and non-local transport models.

\subsection{Outline}
This paper is organized as follows. In section~\ref{sec:model}, we review the equations of motion for a spherical firebrand, along with the characterization of the surrounding wind field. Section~\ref{sec:quant} compares the methods of Monte Carlo simulation, importance sampling, and large deviation theory in approximating the tail end of the firebrand landing distribution.
In section~\ref{sec:lmd}, we define the relative landed mass and derive a formula relating its distribution to the landing distribution. Section~\ref{sec:results} examines the numerical results found by using each method, along with a discussion of their computational cost. We conclude and summarize our results in section~\ref{sec:conclusion}.

\section{Spotting model}
\label{sec:model}
\subsection{Firebrand transport model}
The motion of a firebrand is governed by the interaction of gravitational and aerodynamic forces acting on it.
To describe this motion, we denote the position of a firebrand at time $t$ by $\vc x(t)=(x(t),z(t))$ and its velocity by $\vc v(t) = \dot{\vc x}(t)$. We denote the wind velocity field by $\vc u(\vc x,t)$. 
For simplicity, here we assume that the firebrand moves in a two-dimensional plane, but the transport model is also valid in three dimensions.
Then the equations of motion for a spherical firebrand are given by~\cite{maxey1983,kim1998, babiano2000, linn2007,MR_EUR,langlois2015},
\begin{align}
	\label{eq:fullmotion}
	m(t) \dot{\vc v} - \dot{m}(t) \vc v_{rel}  =& \frac{1}{2} \rho_{f} A_c C_d | \vc u - \vc v |(\vc u - \vc v) && \text{(Quadratic drag)} \nonumber  \\
	&- (m(t) - \rho_{f}V) \vc g && \text{(Gravity \& Buoyancy)} \nonumber \\
	&+ \rho_{f} V \frac{D \vc u}{D t} && \text{\parbox{3cm}{(Pressure gradient)}} \nonumber \\
	&- \frac{1}{2} \rho_{f} V (\dot{\vc v} - \frac{D \vc u}{D  t}). && \text{(Added mass)} 
\end{align}
We denote the fluid density by $\rho_f$, the drag coefficient by $C_d$, cross sectional area of the firebrand by $A_c$, 
firebrand mass by $m(t)$, and firebrand volume by $V=4\pi r^3/3$ where $r$ denotes the firebrand radius.  Table~\ref{tab:table} contains all parameters, their units, and numerical values used here. The right-hand side represents various forces exerted on the firebrand. The first term represents the empirical law of quadratic drag force. The second term represents gravitational and buoyancy forces. The third term accounts for the pressure gradient exerted by the undisturbed fluid. Finally, the fourth term is the added mass effect as  a result of the acceleration of the firebrand with respect to the fluid.

The left-hand side of equation~\eqref{eq:fullmotion} represents the rate of change of momentum for a combusting firebrand~\cite{sommerfeld1952,plastino1992}. The vector $\vc v_{rel}(t)$ represents the velocity of matter leaving the firebrand relative to its center of mass. Following~\cite{manzello2020}, we assume that, as the firebrand combusts, burnt matter leaves it isotropically in all directions. Therefore, it is reasonable to assume that $\vc v_{rel}(t) = 0$ for all time $t$. The fluid density $\rho_f$ is typically much smaller than the firebrand density $\rho_p$, which implies that the buoyancy, pressure gradient, and added mass terms are negligible.

The simplified equations of motion are then given by
\begin{equation}
	\label{eq:motion}
	\dot{\vc x} = \vc v,\quad m(t) \dot{\vc v} = \frac{1}{2}\rho_f A_c C_d |\vc{u}(\vc{x},t)-\vc{v}| (\vc{u}(\vc{x},t)-\vc{v}) - m \left (t \right) \vc{g},
\end{equation}
supplied with the initial conditions $\vc x(0) = \vc x_0$ and $\vc v(0) = \vc v_0$. Here, $\vc g = (0,1)^\top$ where $g=9.8$ $\mbox m/\mbox s^2$ is the constant gravitational acceleration.
As the firebrands are lofted into the air, the create a non-localized distribution of initial positions $\vc x_0$ and
velocities $\vc v_0$. This distribution depends on the height of the canopy, the convective plume, and the size of the firebrand. However, following Bhutia et al.~\cite{bhutia2010}, we make the simplifying assumption that firebrand transport begins from a point source located $50$ meters above the origin, and the firebrand is initially motionless. This corresponds to initial position $\vc x_0 = (0,50)$ and initial velocity $\vc v_0 = (0,0)$.

Of course, the wind velocity field $\vc u \left ( \vc x , t
\right )$ in equation~\eqref{eq:motion} also needs to be supplied. In CFD packages such as HIGRAD/FIRETEC, the wind is obtained by solving the relevant Navier--Stokes equation.
This constitutes the most computationally expensive part of firebrand trajectory computation, and by extension, the most expensive part of spotting distance estimation.  To avoid this computational cost and focus our attention on quantifying the spotting distribution, we use analytically prescribed wind fields. A common choice in firebrand research is a logarithmic wind profile~\cite{tse1998,bhutia2010,jenkins2009,Hillen2016}, given by
\begin{equation}
	\label{eq:logwind}
	\vc u(\vc{x}) := \begin{pmatrix}
		\frac{v_*}{\kappa}\ln(\frac{z}{\epsilon}) \\ 0
	\end{pmatrix},
\end{equation}
where $v_*$ denotes the friction velocity, $\kappa$ is the von Karman's constant, and $\epsilon$ denotes surface roughness length scale, describing the surface vegetation height. In addition to this logarithmic velocity, we also use the bounded wind field,
\begin{equation}
	\label{eq:tanhwind}
	\vc u(\vc{x}) := \begin{pmatrix}
		U \left (1-\left ( \tanh \left (\frac{z}{H} \right) - 1 \right)^2 \right ) \\ 0
	\end{pmatrix}.
\end{equation}
The horizontal component of this velocity field approaches the free stream velocity $U$ as the height $z$ approaches the boundary layer thickness $H$. For $z>H$, the horizontal wind speed remains approximately constant at $U$. We report all our results for both logarithmic and hyperbolic tangent profiles described above.
The numerical value of all parameters are reported in Table~\ref{tab:table}.
\begin{table}
	\centering
	$$
	\begin{array}{|ccrl|}
		\hline
		\text{Variable} & \text{Physical meaning} & \text{Value}  & \text{Units} \\
		\hline
		\rho_f & \text{Fluid density} & 1.204  & kg/m^3 \\
		A_c & \text{Cross sectional area} & -  & m^2 \\
		C_d & \text{Drag coefficient} & 0.45 & - \\
		V & \text{Volume}  & - & m^3 \\
		U & \text{Boundary layer velocity} & 5 & m/s \\
		H & \text{Boundary layer height} & 25 & m \\
		v_* & \text{Friction velocity} & 0.7 & m/s \\
		\kappa & \text{von Karman's constant} & 0.4 & - \\
		\epsilon & \text{Surface vegetation height} & 0.05 & m \\
		\eta & \text{Combustion constant}& 2.86 \times 10^{-4} & s^{-2} \\
		\rho_p & \text{firebrand density} & 513 & kg/m^3 \\
		m(t) & \text{Firebrand mass} & - & kg \\
		p(r) & \text{Nominal size distribution}& - & -\\
		q(r)  & \text{Proposal size distribution}& - & -\\
		f_L(\ell) & \text{Landing distribution} & - & -\\
		f_M(\ell) & \text{Final mass distribution} & - & -\\
		g(\ell) & \text{Landed mass distribution} & - & -\\
		\hline
	\end{array}
	$$
	\caption{Model parameters and their physical dimensions.}
	\label{tab:table}
\end{table}

\subsection{Combustion model}\label{sec:comb}
While firebrands are in flight, they simultaneously burn, resulting in a time-varying mass $m(t)$. Martin~\cite{martin2013} gives a detailed overview of combustion models of firebrands in flight.

The simplest combustion model assumes that the loss of mass is linear in time,
$m(t) = m_0 - ct$,
where $m_0$ is the initial mass of the firebrand and $c$ is the combustion rate. This model has a major flaw in that the mass of the firebrand becomes negative for a large enough value of time $t$.  Tarifa et al.~\cite{Tarifa1963} proposed the empirical model,
\begin{equation}
	\rho_{p}(t) = \frac{\rho_{p}(0)}{1+ \eta t^2},
	\label{eq:C1}
\end{equation}
where the firebrand density at time $t$ is given by $\rho_{p}(t)$ and $\eta = 2.86 \times 10^{-4}$ is a constant determined by analyzing firebrand experiments.
Assuming that the shape of the firebrand remains the same during combustion~\cite{martin2013}, equation~\eqref{eq:C1}
can be written in the equivalent form,
\begin{equation}
	m(t) = \frac{m(0)}{1+ \eta t^2},
	\label{eq:C2}
\end{equation}
using the fact that $m(t) = V\rho_p(t)$ where  $V$ denotes the firebrand volume.

There exist more complex models for combustion.
For instance, Tse and Fernandez-Pello~\cite{tse1998} developed a model that used Nusselt's shrinking drop theory to compute the change in particle diameter of a burning firebrand. Their model is in agreement with the experimental results of Tarifa et al~\cite{Tarifa1963}.

Albini~\cite{albini1979} used a model that assumed that the mass loss rate due to combustion is proportional to the rate of the supply of air to the surface of the firebrand. In this paper, we use the empirical model~\eqref{eq:C2}, although more complex combustion models can be used with no significant change to the methods introduced in section~\ref{sec:quant}.

\subsection{Assumptions}\label{sec:assump}
We made a number of simplifying assumptions in sections~\ref{sec:model} and~\ref{sec:comb}. Most of these assumptions are justified and do not significantly alter the results. For instance, we neglected the buoyancy, added mass and pressure gradient effects in equation~\eqref{eq:fullmotion} based on the fact that the density of the firebrand is much larger than the fluid density.
This assumption is routine and it is justified since inclusion of the neglected forces does not significantly change the firebrand trajectories.

However, there are some crucial assumptions made to reduce the computational cost. Although, these assumptions are made for this first study of rare events in spotting, they should ultimately be relaxed in future studies. We list these crucial assumptions below.
\begin{enumerate}
\item Point sources: We assume that the firebrands are released from a point source at $\vc x_0$ and with the deterministic velocity $\vc v_0$. In truth, the initial position and velocity of the firebrands themselves are random variables.
\item Spherical firebrands: We assume that all firebrands are spherical with variable radius size. In reality, firebrands come in various complex shapes. Earlier studies have focused primarily on spherical, cylindrical, and disk-shaped firebrands~\cite{linn2007}.
\item Prescribed steady two-dimensional wind: We assumed that the firebrands move in a two-dimensional plane aligned with the predominant direction of the wind. Furthermore, we consider two prescribed steady and laminar velocity fields. In reality,
the wind velocity is turbulent and three-dimensional, occurring in areas with obstacles and complex topography.
\end{enumerate}
Although we make the above simplifying assumptions, the methodology and our main findings are applicable to more complex flows.

\section{Rare event quantification}
\label{sec:quant}
In this section, we review three methods for quantifying the spotting distribution with a special focus on the 
tail of the distribution where the firebrands land farthest from the original fire.
One of the main factors that determines the landing distance is the size of the firebrand, i.e., the radius $r$ of a spherical firebrand. We treat this radius as a random variable $R$ which is distributed according to a known probability density $p(r)$. 

Given firebrands whose radii are a random variable, our goal is to determine the probability distribution of the landing locations and to find the distribution of the relative landed mass as a function of space. 

\subsection{Monte Carlo method}
\label{sec:mc}
The most straightforward method for estimating the spotting distribution is the crude Monte Carlo (MC) method. To describe this method, we first define the map $L: \mathbb{R} \to \mathbb{R}$ that maps the firebrand radius $R$ to a landing distance $L(R)$ obtained by using equation~\eqref{eq:motion} to advect firebrands. Note that $L(R) = x(t_\ast)$ for a firebrand with initial mass $m(0) = (4\pi R^3/3)\rho_p$, where $t_\ast$ is the time it takes for the firebrand to land, so that $z(t_\ast) =0$.

Given the probability distribution of the radii $R$, we want to estimate the probability that a firebrand lands at a distance
$\ell$ from the source. More precisely, consider the interval $D(\ell,\Delta \ell) := [\ell - \Delta \ell/2, \ell + \Delta \ell/2)$ centered at $\ell$ with a small length $\Delta \ell$. We would like to estimate the probability $\mathbb P(L(R)\in D(\ell,\Delta \ell))$, which is the probability that a firebrand lands in the interval $D(\ell,\Delta \ell)$. In the following, we use the shorthand $D(\ell)$ in place of $D(\ell,\Delta \ell)$.
If $f_L$ is the probability density function (PDF) associated with the random variable $L(R)$, we have 
\begin{equation}
\label{eq:pdfapprox}
f_L(\ell) \simeq \frac{\mathbb{P}(L(R)\in D(\ell))}{\Delta \ell}.
\end{equation}

We express the probability in terms of the integral,
\begin{align}
\mathbb{P}(L(R) \in D(\ell)) &= \int_{L^{-1}\left(D \left ( \ell \right)\right)} p(r)\id r \nonumber\\
&= \int_{0}^{\infty} 	\mathbbm{1}_{\ell}\left(L(r)\right) p(r)\id r,
\label{eq:prob_int}
\end{align}
where $\mathbbm{1}_{\ell}(\cdot)$ is shorthand for the indicator function of the set $D(\ell)$,
\begin{equation*}
\mathbbm{1}_{\ell}(\hat{\ell}) := 
\begin{cases}
	1,&  \hat{\ell} \in D(\ell), \\
	0,&  \textrm{otherwise}.
\end{cases}
\end{equation*}

The MC method is a straightforward method for estimating integral~\eqref{eq:prob_int}.
Consider $N$ independent, identically distributed (i.i.d.) realizations of the firebrand radii, denoted by $R_i$ for $i = 1, \cdots, N$. The corresponding landing distances are given by $L_i := L(R_i)$. We can approximate~\eqref{eq:prob_int} with the MC estimator,
\begin{equation}
\label{eq:MCapprox}
P_{MC}(\ell) := \frac{1}{N} \sum_{i=1}^N \mathbbm{1}_{\ell}\left(L_i\right)\simeq \mathbb{P}(L(R) \in D(\ell)),
\end{equation}
where radius realizations $R_i$ are drawn from the distribution $p$ and the firebrands are advected using equation~\eqref{eq:motion} to obtain their corresponding landing distances $L_{i}$. The MC estimator $P_{MC}$ computes the ratio of the firebrands that land in the interval $D(\ell)$ to the total number of firebrands.
\begin{algorithm}[t]
\DontPrintSemicolon
\SetAlgoLined
\textbf{Inputs}: Sample size $N_{MC}$ and number of intervals $K$.\;
\For{$i = 1, \cdots, N_{MC}$}{
	Generate firebrand sizes $R_i$ from distribution $p$.\;
	Advect each particle using equation~\eqref{eq:motion} to obtain the corresponding landing distances $L_i$.}
Divide the landing interval $[0,\max_i L_i]$ into $K$ equisized intervals $D(\ell_j)$ with centers $\{\ell_j \}_{j=1}^{K}$
and widths $\Delta\ell$.\;
\For{$j = 1, \cdots, K$}{
	For landing distance of interest $\ell_j$, use equation~\eqref{eq:MCapprox} to approximate $\mathbb{P}(L(R)\in D(\ell_j))$.\;
	Estimate the landing PDF $f_L(\ell_{j})$ using approximation~\eqref{eq:pdfapprox}.}
\textbf{Output}: Landing distribution $f_L$
\caption{MC approximation\label{alg:MC}}
\end{algorithm}

As seen in its implementation in Algorithm~\ref{alg:MC}, direct MC simulation is straightforward. A sample of $N_{MC}$ firebrands with radii $R_i$ are drawn from the distribution $p(r)$. Each firebrand is evolved separately under equation~\eqref{eq:motion} to obtain its corresponding landing location $L_i=L(R_i)$. 
Then the landing  interval is divided into $K$ bins with width $\Delta \ell$. Finally, equations~\eqref{eq:pdfapprox} and~\eqref{eq:MCapprox} are used to estimate the landing distribution $f_L$.

The MC estimator is unbiased in the sense that $\mathbb E_p[P_{MC}] = \mathbb P(L(R)\in D(\ell))$, where $\mathbb E_p$ denotes the expected value taken with respect to the probability density $p$.
It is also straightforward to show that the variance of $\sigma^2$ of the estimator $P_{MC}$ is given by 
$P_\ell(1-P_\ell)/N_{MC}$~\cite{bee2009}, where we denoted $\mathbb P(L(R)\in D(\ell))$ by $P_\ell$. The relative error, defined as the ratio of the standard deviation to the mean, is given by $\sigma/P_\ell = \sqrt{(1-P_\ell) /N_{MC}P_\ell}$.
For rare events where the probability $P_\ell$ is very small, the relative error is approximately $\sigma/P_\ell\simeq 1/\sqrt{N_{MC}P_\ell}$. In order to obtain a small relative error, one needs to use an exceedingly large sample size $N_{MC}$. 
For instance, if the probability is $10^{-6}$ one needs a sample size of $100$ millions to obtain a $10\%$ relative error.

As a result, MC method is not practical for quantifying rare spotting events, i.e., spot fires forming far away from the primary fire. Importance sampling, as reviewed in section~\ref{sec:is}, seeks to alleviate this computational cost.

\subsection{Importance sampling}\label{sec:is}
Importance sampling (IS) is a variance reduction method~\cite{kloek1978}. The basic idea behind IS is to modify the sampling distribution so that more samples are obtained from the low probability tail of the landing distribution $f_L$. More precisely, we draw samples from a \emph{proposal distribution} $q$ instead of the nominal distribution $p$ of firebrand radii. We seek the proposal distribution which minimizes the variance in estimating $\mathbb P(L(R)\in D(\ell))$.

Before specifying the optimal distribution $q$, note that 
\begin{align*}
\mathbb{P}(L(R)\in D(\ell)) 
&= \int_{0}^{\infty} \mathbbm{1}_{\ell}(L(r)) p(r)\id r \\
&= \int_{0}^{\infty} \frac{\mathbbm{1}_{\ell}(L(r))p(r)}{q(r)} q(r)\id r,
\end{align*}
for any proposal distribution $q$. 
Of course, for the integrals to be well-defined, we must have $\mathbbm{1}_{\ell}(L(r))p(r) = 0$ when $q(r) = 0$. 
Then the importance sampling estimator is 
\begin{equation}
\label{eq:ISapprox}
P_{IS}(\ell) := \frac{1}{N} \sum_{i=1}^N \frac{	\mathbbm{1}_{\ell}(L_i)p(R_i)}{q(R_i)},
\end{equation}
where radius random variables $R_i$ are now drawn from the proposal distribution $q$.
As in the MC case, the new quantity $P_{IS}$ is  an unbiased estimator since $\mathbb E_q[P_{IS}]=\mathbb{P}(L(R)\in D(\ell))$, where $\mathbb E_q$ is the expected value with respect to the proposal distribution.

The question remains on how to choose the optimal proposal distribution $q$ which minimizes the variance of the estimator $\var [P_{IS}]$. In general, determining this optimal distribution is laborious. However, if we restrict the admissible class of proposal distributions $q$ to the same type of distribution as the nominal distribution $p$, the optimal distribution can be identified more easily~\cite{bee2009,mcleish2015}.
\begin{algorithm}[t]
\DontPrintSemicolon
\SetAlgoLined
\textbf{Inputs}: Sample sizes $N_{IS}$ and $\hat N_{IS}$, and number of intervals $K$.\;
\For{$i = 1, \cdots, N_{IS}$}{
	Generate firebrand sizes $R_i$ from distribution $p$.\;
	Advect each particle using equation~\eqref{eq:motion} to obtain corresponding landing distances $L_i$.}\
Divide the landing interval $[0,\max_i L_i]$ into $K$ equisized intervals $D(\ell_j)$ with centers $\{\ell_j \}_{j=1}^{K}$ and widths $\Delta\ell$.\;
\For{$j = 1, \cdots, K$}{
	For landing distances of interest ${ \ell_{j}}$ solve problem~\eqref{eq:varopt} to obtain $q$.
	\For{$k= 1, \cdots, \hat N_{IS}$}{Generate firebrand sizes $R_k$ from distribution $q$.\;
		Advect each particle using equation~\eqref{eq:motion} to obtain corresponding landing distances $L_k$.}\
	Calculate~\eqref{eq:ISapprox} to approximate $\mathbb{P}(\ell_j - \Delta \ell/2 \leq L(R) \leq \ell_j + \Delta \ell/2)$.\;
	Estimate the landing distribution $f_L(\ell_j)$ using the approximation~\eqref{eq:pdfapprox}.
}
\textbf{Output}: Landing distribution $f_L$
\caption{IS approximation\label{alg:IS}}
\end{algorithm}

More precisely, let $q$ be a lognormal distribution with mean $\mu_q$ and variance $\sigma_q^2$, which are potentially different from the mean $\mu_p$ and variance $\sigma_p^2$ of the firebrand radius distribution $p$.
We would like to determine $\theta := \{ \mu_q, \sigma_q \}$ such that the variance of the estimator $\var [P_{IS}]$ is minimized. The problem is that $\var [P_{IS}]$  is a priori unknown. To rectify this issue, we estimate this variance by running a relatively small MC simulation to compute
\begin{equation}
\label{eq:est_var}
V(\vc{\theta};\ell) := \frac{1}{N_{IS}}\sum_{i=1}^{N_{IS}} \mathbbm{1}_{\ell}(L_i)\frac{p(R_i)}{q(R_i ; \vc{\theta})}\simeq \var[P_{IS}],
\end{equation}
where $R_i\sim p$ and $N_{IS}$ is the MC sample size. Note that unlike the mean estimator~\eqref{eq:ISapprox}, $V(\theta)$ is computed by sampling the radii $R_i$ from the distribution $p$ not the proposal distribution $q$. Also note that the \emph{likelihood ratio} is given by
\begin{equation}
\frac{p(r)}{q(r;\theta)} 
= \frac{\sigma_q}{\sigma_p} \exp \Big \{ -\frac{1}{2} \Big ( \frac{(\ln \left (r\right )-\mu_p)^2}{\sigma_p^2} \Big ) + \frac{1}{2} \Big ( \frac{(\ln \left (r\right )-\mu_q)^2}{\sigma_q^2} \Big ) \Big \},
\end{equation}
where $\mu_p$ and $\sigma_p$ are the known mean and variance of the firebrand size distribution $p$, respectively.

Then, we solve the optimization problem,
\begin{equation}
\label{eq:varopt}
\theta_\ast(\ell):=\arg\min_{\vc{\theta}} V(\vc{\theta};\ell),
\end{equation}
for each $\ell$ to obtain the corresponding optimal proposal distribution $q(r;\theta_\ast)$. Finally, we draw
$\hat N_{IS}$ samples from this optimal distribution in order to compute the importance sampling estimate~\eqref{eq:ISapprox}. Algorithm~\ref{alg:IS} summarizes the entire IS method.

Note that since the optimal $\theta_\ast$ depends on $\ell$, the optimization problem~\eqref{eq:varopt}
must be solved for each spatial interval $D(\ell)$. If there are $K$ intervals, the IS algorithm requires $N_{IS}+K\hat N_{IS}$ samples. 
Since the variance of the estimator $\var [P_{IS}]$ is reduced compared to the crude MC method, accurate approximations can be obtained with even small sample sizes $N_{IS}$ and $\hat N_{IS}$. We show this with numerical examples in section~\ref{sec:results}.

\subsection{Large deviation theory}
\label{sec:ldt}
Recall that estimating the landing distribution $f_L(\ell)$ is most demanding for rare events where $\ell$ is large, or theoretically when $\ell\to\infty$. Large deviation theory refers to a collection of methods that focus on this asymptotic limit of probability distributions~\cite{varadhan1984,hollander2000,dembo1998,bucklew2004}. An LDT method that is best suited for application to spotting was only recently developed by Dematteis et al.~\cite{Dematteis2019} (also see~\cite{Dematteis2018,tong2021}). The theory is quite technical and therefore here we only review its essential aspects and formulate it for its application to spotting.

Unlike crude MC and importance sampling, LDT does not rely on sampling. Instead, LDT provides an asymptotic expression for evaluating the probability $\mathbb P(L(R)\geq \ell)$ for large values of $\ell$. Evaluating the LDT estimate requires solving an optimization problem, but not sampling.

The LDT theory is best described for Gaussian random variables. Therefore, we define log radius $Z:=\ln R$ which is a Gaussian random variable since the firebrand radius $R$ is lognormal. The landing distance for a firebrand with log radius $Z$ is given by  $L(\exp(Z))$ which, for notational simplicity, we denote by $L(Z)$.
LDT predicts that, for large enough $\ell$, the probability $\mathbb P(L(Z)\geq \ell)$ is approximately given by
\begin{equation}
\label{eq:approx}
P_{LD}(\ell) := (2\pi)^{-1/2} \frac{1}{\sqrt{2I(Z^*(\ell))}} \exp \left ( -I(Z^{*}(\ell))\right)\simeq \mathbb P(L(Z)\geq \ell).
\end{equation}
where $I:\mathbb R\to\mathbb R$ is the so-called rate function, 
\begin{equation}
I(z) := \max_{\eta \in \mathbb{R}} [\eta z - \ln T(\eta)],
\end{equation}
and $T(\eta)=\mathbb{E} \left[\exp(\eta Z) \right]$ is the moment generating function for the Gaussian random variable $Z$.
The log radius $Z^\ast$ in equation~\eqref{eq:approx} is the solution to the optimization problem,
\begin{equation}
\label{eq:constrained}
Z^*(\ell) := \arg \min_{Z \in \Omega(\ell)}I(Z),
\end{equation}
where $\Omega(\ell) := \{ Z \in \mathbb{R} :L(Z) \geq \ell \}$ is the set of all log radii such that the corresponding landing distance $L(z)$ exceeds $\ell$.

A few remarks are in order here. First, since $Z$ has a Gaussian distribution, the rate function $I(Z)$ can be computed explicitly. Note that the moment generating function for a Gaussian random variable is given by $T(\eta) =\mathbb{E} \left[ \exp \left( \eta Z \right) \right] = \exp \left ( \eta \mu_0 + \sigma_0^2 \eta^2/2 \right)$ where $\mu_0$ and $\sigma_0^2$ are the mean and variance of $Z$, respectively. As a result, the rate function becomes
\begin{equation}
I(z) = \frac{1}{2\sigma_0^2}(z- \mu_0)^2.
\end{equation}
Therefore, the only optimization required is when finding $Z^*(\ell)$, which involves solving the constrained optimization problem~\eqref{eq:constrained}. But, as shown by Tong et al.~\cite{tong2021}, the optimizer lies on the boundary of $\Omega(\ell)$. Therefore, equation~\eqref{eq:constrained} can be rewritten as the unconstrained optimization problem,
\begin{equation}
\label{eq:unconstrained}
Z^*(\lambda) := \arg \min_{Z \in \mathbb{R}} \left[ I(Z) - \lambda L(Z)\right],
\end{equation}
with the Lagrange multiplier $\lambda > 0$. 

In practice, we choose a sequence of $N_\lambda$ Lagrange multipliers $0<\lambda_1<\lambda_2<\cdots<\lambda_{N_\lambda}$ and, for each $\lambda_i$, solve optimization~\eqref{eq:unconstrained}. This then determines a corresponding sequence of log radii $Z^\ast(\lambda_i)$ and landing distances $\ell_i = L(Z^\ast(\lambda_i))$.  Since the optimizer $Z^\ast(\ell_i)$ of~\eqref{eq:constrained} lies on the boundary of $\Omega(\ell_i)$, it coincides with the optimizer $Z^\ast(\lambda_i)$ of~\eqref{eq:unconstrained} with $\ell_i = L(Z^\ast(\lambda_i))$.
Larger values of $\lambda_i$ correspond to more extreme landing locations.

With the optimizers $Z^\ast(\ell_i)$ at hand, the LDT approximation $P_{LD}(\ell_i)$ can be computed from~\eqref{eq:approx}.
Note that $P_{LD}(\ell_i)- P_{LD}(\ell_{i+1})$ estimates the probability that a firebrand lands in the interval $[\ell_i,\ell_{i+1})$.
Therefore, the probability density $f_L(\ell_i)$ can be approximated by
\begin{equation}
f_L(\ell_i) \simeq \frac{P_{LD}(\ell_i)-P_{LD}(\ell_{i+1})}{\ell_{i+1}-\ell_{i}},
\end{equation}
as long as $\ell_{i+1}-\ell_i$, or equivalently $\lambda_{i+1}-\lambda_i$, is sufficiently small. Note that $P_{LD}(\ell_i)\geq P_{LD}(\ell_{i+1})$ since $\ell_{i+1}\geq \ell_i$. The entire LDT approximation is summarized in Algorithm~\ref{alg:LDT}.

\begin{algorithm}[t]
\DontPrintSemicolon
\SetAlgoLined
\textbf{Inputs}: Positive increasing sequence $\lambda_1<\lambda_2<\cdots<\lambda_{N_\lambda}$.\
\\
\For{$i = 1, \cdots, N_\lambda$}{
	Solve optimization problem~\eqref{eq:unconstrained} with $\lambda=\lambda_i$.\
	\\
	Advect a firebrand of size $Z^*(\lambda_i)$ to obtain the landing distance $\ell_i = L(Z^*(\lambda_i))$.\
	\\
	Calculate $P_{LD}(\ell_i)$ using equation~\eqref{eq:approx} with $Z^\ast(\ell_i)  = Z^\ast(\lambda_i)$.}\
\For{$i = 1, \cdots, N_\lambda-1$}{
	Estimate the landing PDF at landing distance $\ell_i$ using
	$$
	f_L(\ell_i) \simeq \frac{P_{LD}(\ell_i)-P_{LD}(\ell_{i+1})}{\ell_{i+1}-\ell_{i}}.
	$$
}
\textbf{Output}: Landing distribution $f_L$
\caption{LDT approximation\label{alg:LDT}}
\end{algorithm}

We recall that the LDT method does not require sampling and thus it is computationally less expensive that Monte Carlo and importance sampling methods.
However, this speed up comes at a cost. First, the LDT approximation~\eqref{eq:approx} is only valid asymptotically, i.e., for large enough $\ell$.
As a result, LDT can only be applied to quantifying spotting probabilities at large distances. Furthermore, we are unaware of any estimates on the accuracy of the LDT approximation, e.g., the variance of the LDT approximation. 

Nonetheless, as we show in section~\ref{sec:results}, in our numerical experiments LDT approximation agrees very well with the more costly MC and IS estimates. Furthermore, in our experiments, the range of validity of LDT is not excessively small. In fact, it is accurate for landing locations within one standard deviation from the mode of the distribution.

\section{Relative landed mass distribution}
\label{sec:lmd}
The landing distribution $f_L$, that was estimated in section~\ref{sec:quant}, quantifies the proportion of the firebrands landing at a distance $\ell$,
regardless of their size or mass. However, burning firebrands with larger mass are more likely to start a fire at their landing location. As a result, it is perhaps more relevant to quantify the proportion of firebrand mass landed in a small interval at distance $\ell$ from the main fire. We refer to this quantity as the \emph{relative landed mass distribution} and derive an equation that enables us to compute this quantity from the landing distribution $f_L$.

We denote the probability density associated with the relative landed mass distribution by $g: \mathbb{R} \to \mathbb{R}^+$ so that $g(\ell) \Delta \ell$ estimates the ratio of the mass landed in the interval $D(\ell)$ to the total mass landed anywhere. More precisely, the density $g$ is given by 
\begin{equation}
\label{eq:masspdf}
g(\ell) := \lim_{\Delta\ell\to 0^+}\frac{1}{\Delta \ell} \lim_{N\to \infty} \frac{\sum_{i=1}^{N}\mathbbm{1}_{\ell} \left ( L_i \right ) M_i }{\sum_{i=1}^{N} M_i},
\end{equation}
where $L_i=L(R_i)$ and the firebrand radii $R_i$ are drawn from the probability density $p$. The random variable $M_i$ is the mass of the firebrand with radius $R_i$ at the time of landing, which is computed using the combustion model~\eqref{eq:C2}.

\begin{figure}
\centering
\subfigure[]{\label{fig:philog}\includegraphics[width=0.49\textwidth]{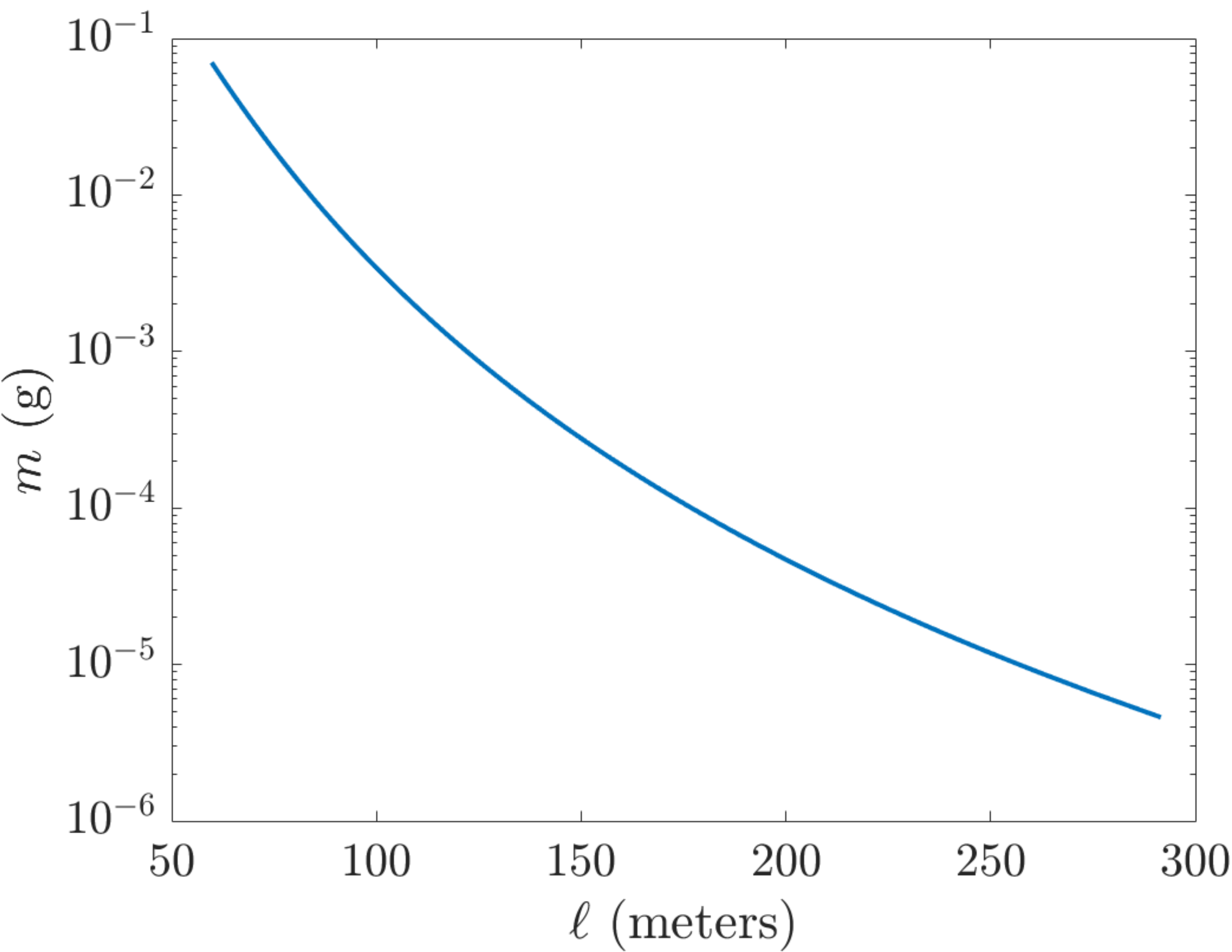}}
\subfigure[]{\label{fig:phitanh}\includegraphics[width=0.49\textwidth]{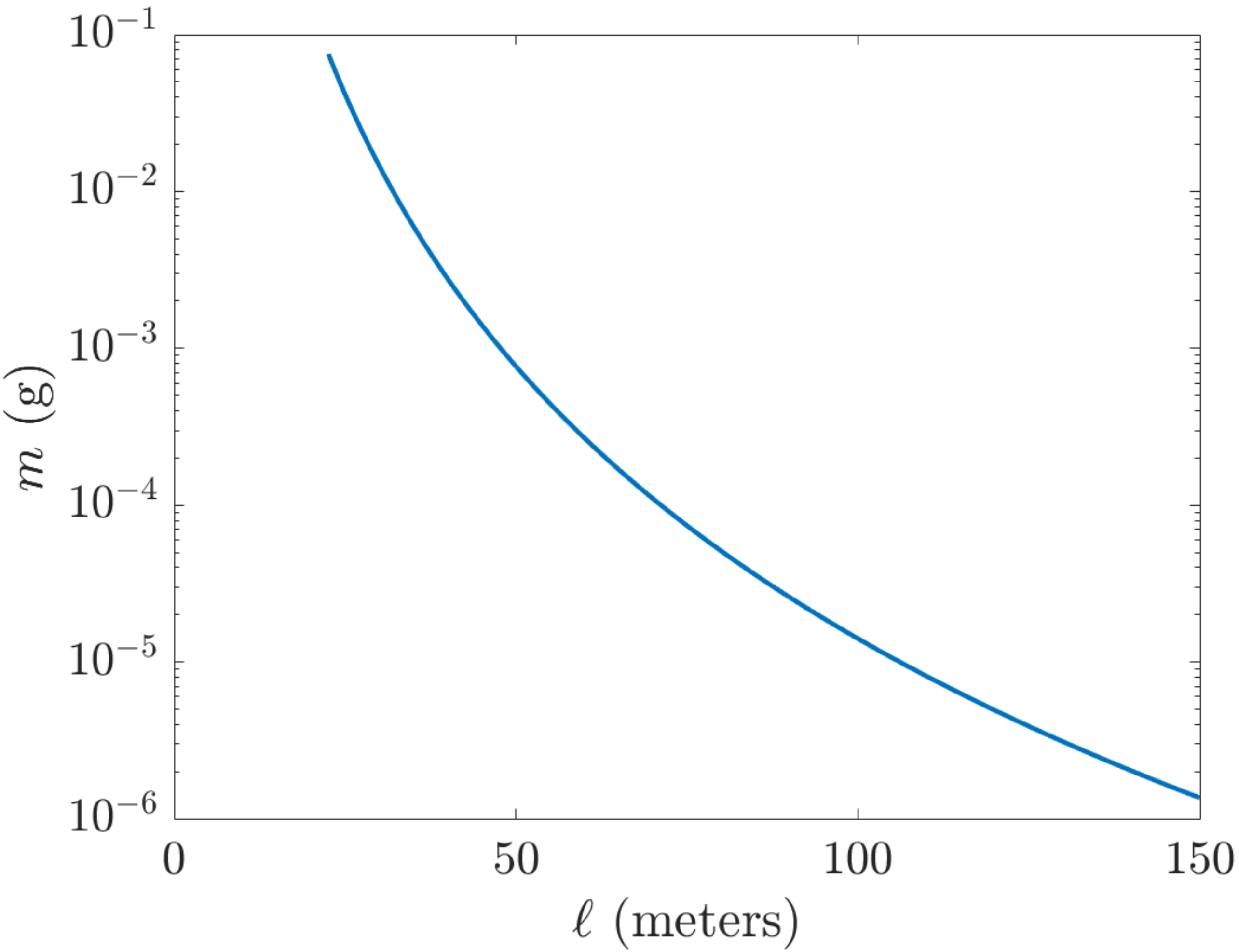}}

\caption{
	Comparison of final mass functions $\phi$ for each wind field. (a) Final mass function $\phi$ corresponding to wind field~\eqref{eq:logwind}. (b) Final mass function $\phi$ corresponding to wind field~\eqref{eq:tanhwind}. 
}
\label{fig:phicomparison}
\end{figure}

We define $\phi : \mathbb{R} \to \mathbb{R}^+$ as the map between the landing distance and its associated landing mass so that $M_i= \phi(L_i)$.
As shown in figure~\ref{fig:phicomparison}, this is a one-to-one decreasing function. This function is monotonically decreasing because the firebrands that land farther are airborne for a longer time and therefore have more time to burn and lose mass.

We note that relative landed mass density $g(\ell)$ should not be confused with the landed mass density $f_M(m)$ of $M_i$. The quantity $f_M(m)\Delta m$ measures the probability that the landed firebrand mass 
falls in the mass interval $[m-\Delta m/2,m+\Delta m/2]$. As such, $f_M$ contains no information about the landing location. In contrast, $g(\ell)$ estimates the proportion of mass landed in the spatial interval $[\ell-\Delta \ell/2,\ell+\Delta\ell/2]$.

In principle, the relative landed mass density~\eqref{eq:masspdf} can be approximated using the MC method by sampling a large number of firebrands, computing their landing locations $L_i$ and the corresponding mass $M_i$ at the landing time. However, as in the case of the landing distribution $f_L$, this estimate will be inaccurate for large $\ell$ unless we use a prohibitively large sample size. To avoid this problem, here we show that 
\begin{equation}
g(\ell) = \frac{\phi(\ell) f_L(\ell)}{\mathbbm{E} \left [M \right ]},
\label{eq:lmd_equiv}
\end{equation}
where $\phi(\ell)$ is the landing mass of firebrands that land at the distance $\ell$ (see figure~\ref{fig:phicomparison}), $f_L$ is the landing distribution computed in section~\ref{sec:quant}, and $\mathbb E[M]$ is the expected value of the landed mass. The remainder of this section is devoted to proving the estimate~\eqref{eq:lmd_equiv}.

We first rewrite equation~\eqref{eq:masspdf} by multiplying $\frac{1}{N}$ in the numerator and the denominator which yields
\begin{equation}
\label{eq:masspdfmod}
g(\ell) =\lim_{\Delta \ell\to0^+} \frac{1}{\Delta \ell}\lim_{N\to\infty} \frac{\frac{1}{N}\sum_{i=1}^{N}\mathbbm{1}_{\ell} \left ( L_i \right ) \phi \left ( L_i \right) }{\frac{1}{N} \sum_{i=1}^{N} \phi \left ( L_i \right)}, \qquad R_i \sim p.
\end{equation}
The denominator of this expression is the expected value of the landed mass, so that
\begin{equation}
\lim_{N\to\infty}\frac{1}{N} \sum_{i=1}^{N} \phi \left ( L_i \right) = \mathbb E[M]:=\int_0^\infty mf_M(m)\id m. 
\label{eq:denom}
\end{equation}
The numerator is the expected value of the mass that lands in the interval $D \left (\ell \right)$, 
\begin{equation}
\lim_{N\to\infty}\frac{1}{N} \sum_{i=1}^{N} \mathbbm{1}_{\ell} (L_i)\phi \left ( L_i \right) =
\int_{\phi \left(D \left(\ell\right) \right)} m f_M(m)\id m.
\label{eq:numer1}
\end{equation}
Recall that by definition $m = \phi \left (\ell \right)$. Therefore, by the change of variable formula, we have
\begin{equation}
\int_{\phi \left(D \left(\ell\right) \right)} m f_M(m)\id m= \int_{D \left(\ell\right)} -\phi(\ell) f_M(\phi(\ell))\phi'(\ell)\id \ell,
\label{eq:numer2}
\end{equation}
where the minus sign is due to the fact that $\phi$ is monotonically decreasing.

On the other hand, the probability that a firebrand lands in $D(\ell)$ is equal to the probability that the firebrand mass at the time of landing is in $\phi(D(\ell))$.
More precisely, we have 

\begin{equation}
\int_{D(\ell)} f_L \left ( \ell \right) \id\ell=\int_{\phi(D(\ell))} f_M \left ( m \right ) \id m = - \int_{D(\ell)} f_M \left ( \phi \left ( \ell \right) \right) \phi' \left ( \ell \right) \id\ell,
\label{eq:fL_fM}
\end{equation}
where we again used change of variables for the last identity. Since~\eqref{eq:fL_fM} must hold for any arbitrary interval $D(\ell)$, we obtain
$f_L \left( \ell \right) = - f_M \left( \phi \left(\ell \right) \right) \phi' \left(\ell \right)$. Substituting this in equation~\eqref{eq:numer2} and using equation~\eqref{eq:numer1}, we obtain
\begin{equation}
\lim_{N\to\infty}\frac{1}{N} \sum_{i=1}^{N} \mathbbm{1}_{\ell} (L_i)\phi \left ( L_i \right) =
\int_{D \left(\ell \right)} \phi(\ell) f_L(\ell)  \id\ell 
\simeq \Delta \ell\, \phi(\ell) f_L(\ell).
\label{eq:numer3}
\end{equation}
Finally, combining equations~\eqref{eq:masspdfmod},~\eqref{eq:denom} and~\eqref{eq:numer3} gives the desired result~\eqref{eq:lmd_equiv}.

\section{Numerical results and discussion}
\label{sec:results}
\begin{figure}
\centering
\includegraphics[width=0.4\textwidth]{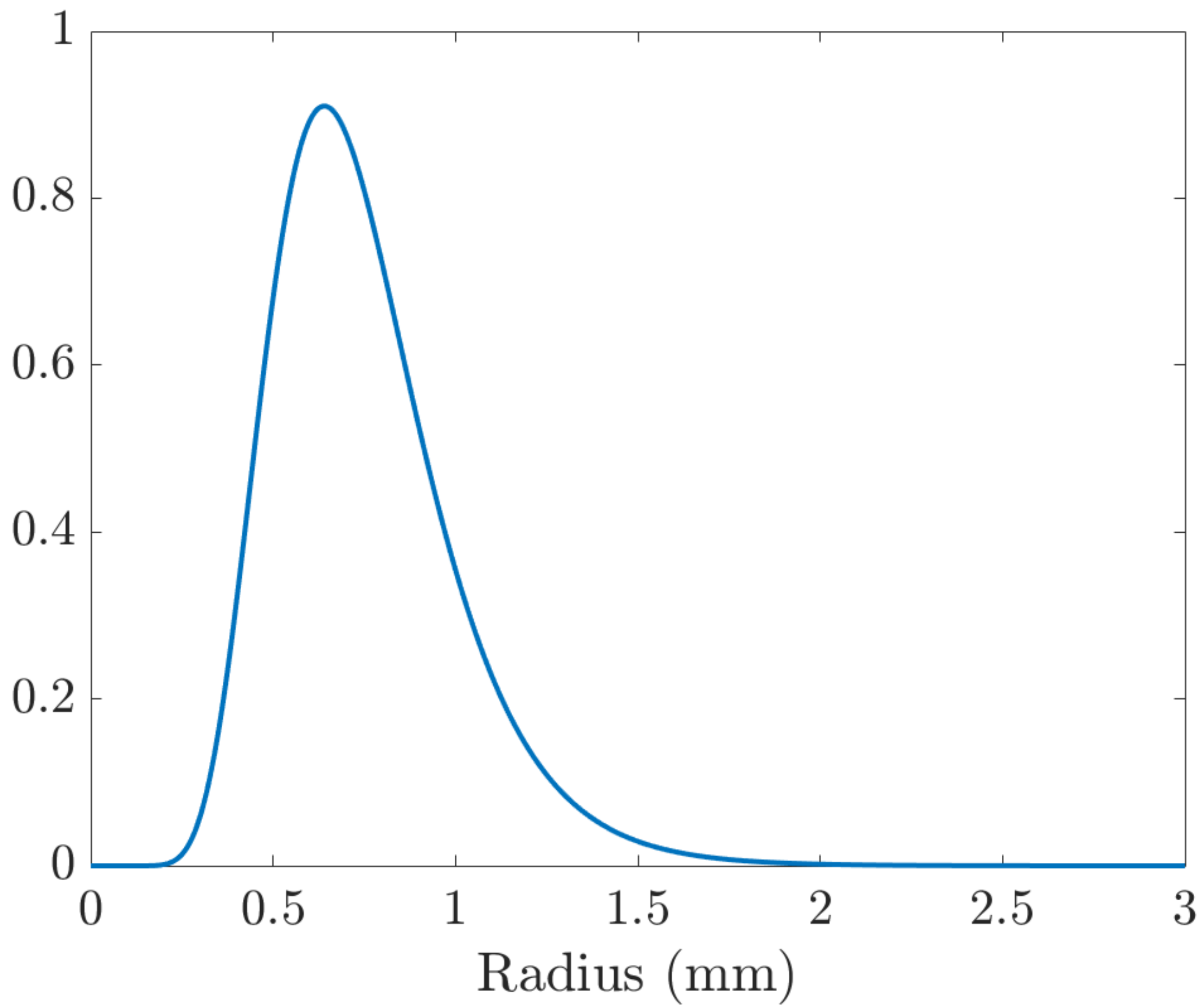}
\caption{Probability density function of a lognormally distributed firebrand radius $R$ with mean $0.75$ millimeters and variance $0.125$ millimeters.}
\label{fig:sizepdf}
\end{figure}

In this section, we report our numerical results, comparing MC, IS, and LDT methods as described in section~\ref{sec:quant}.
We particularly focus on the trade-off between accuracy and computational cost.
We assume that the firebrand radius $R$ is a lognormally distributed random variable, with a mean radius of $0.75$ millimeters and a variance of $0.125$ millimeters. The corresponding probability density function $p$ is shown in figure~\ref{fig:sizepdf}.

\begin{figure}
\centering
\subfigure[]{\label{fig:a}\includegraphics[width=0.32\textwidth]{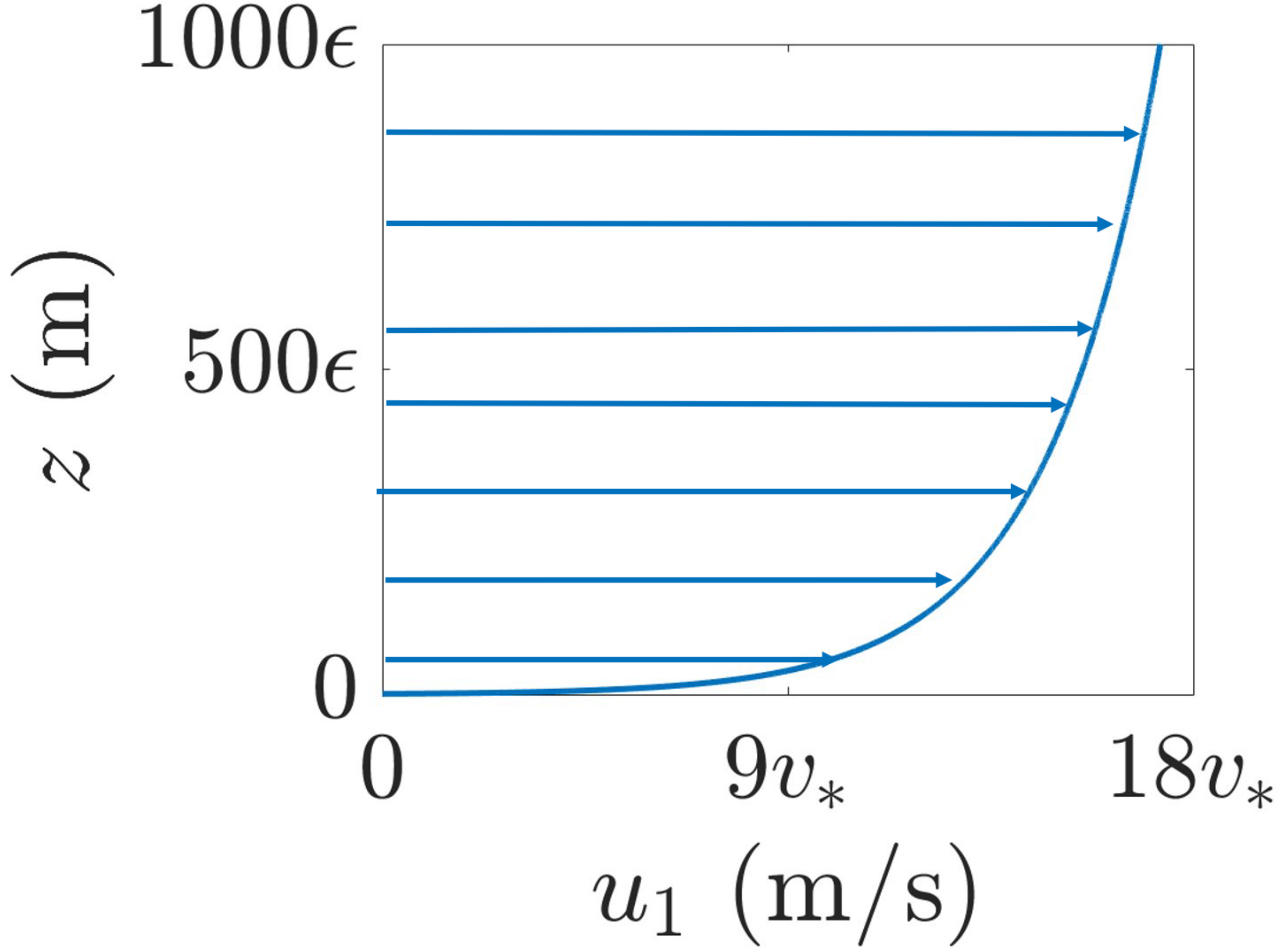}}
\subfigure[]{\label{fig:b}\includegraphics[width=0.32\textwidth]{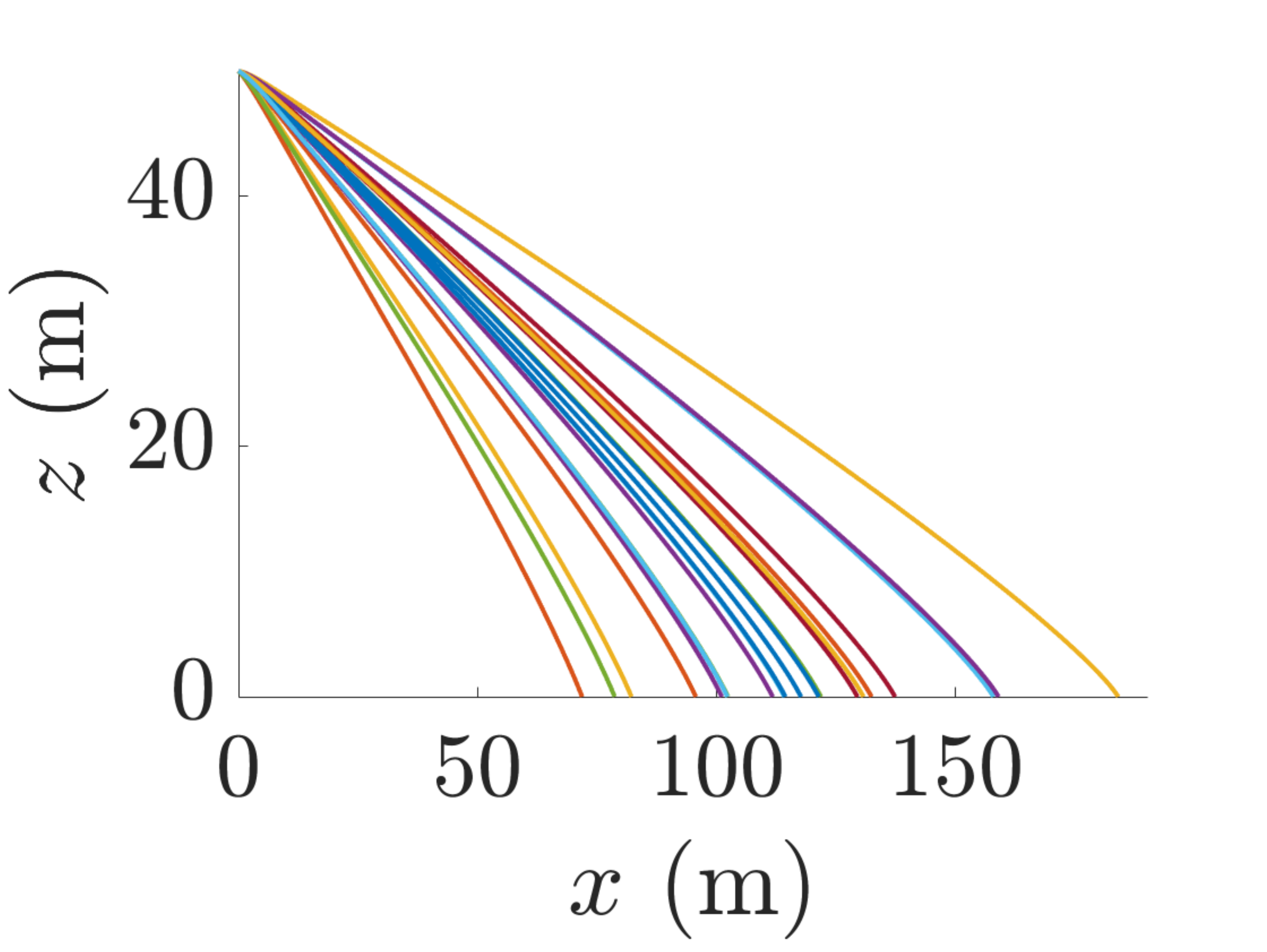}}
\subfigure[]{\label{fig:c}\includegraphics[width=0.32\textwidth]{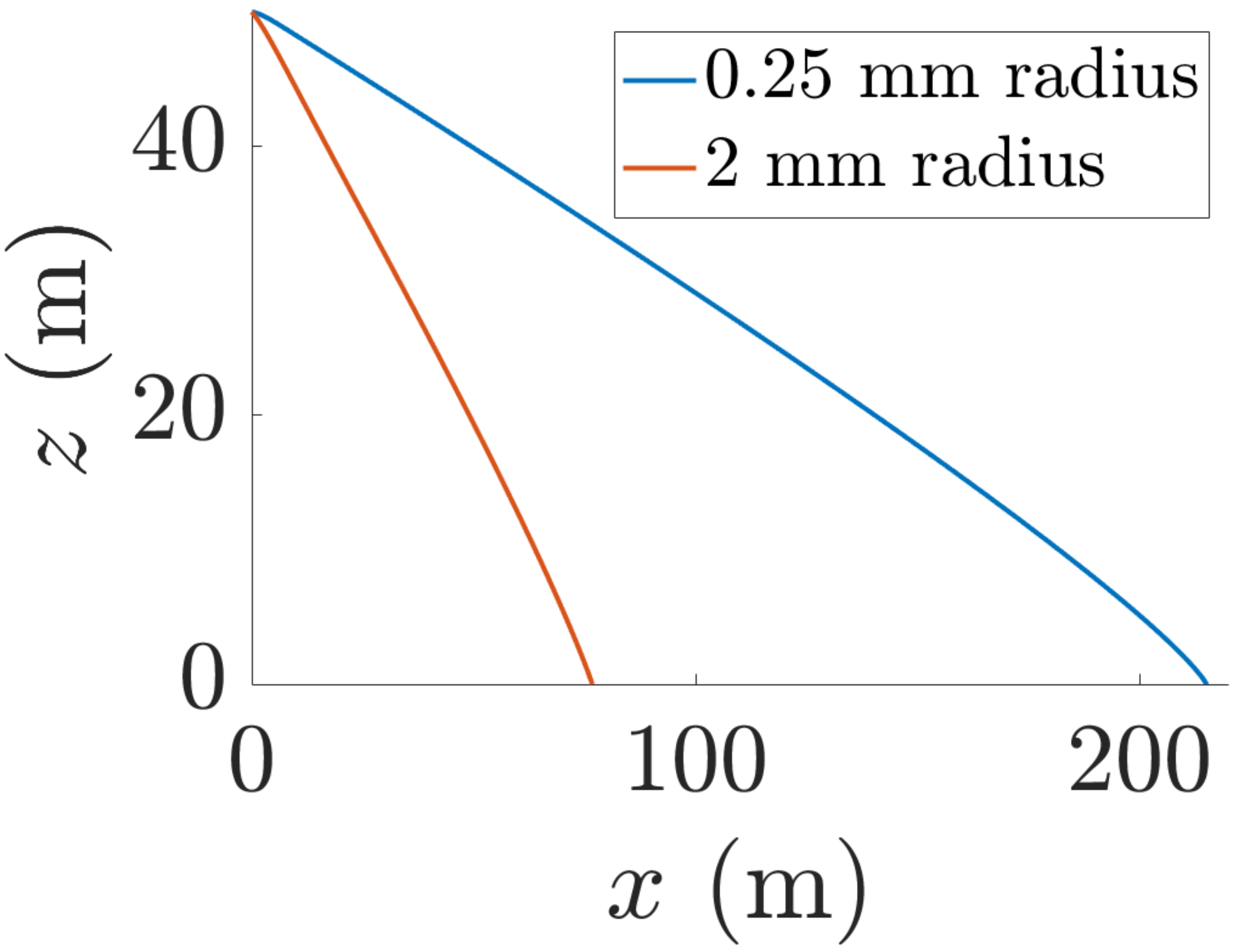}}
\subfigure[]{\label{fig:d}\includegraphics[width=0.32\textwidth]{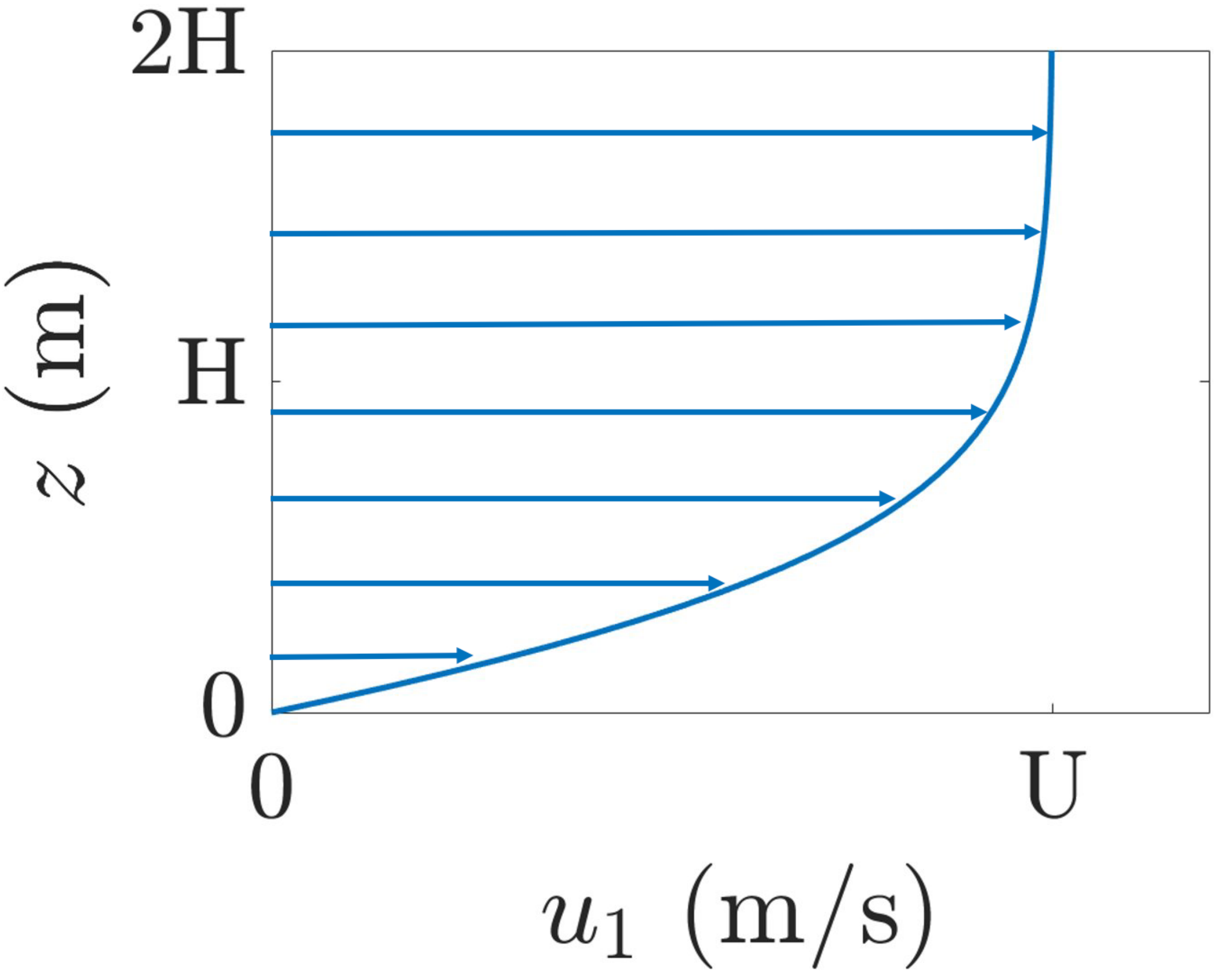}}
\subfigure[]{\label{fig:e}\includegraphics[width=0.32\textwidth]{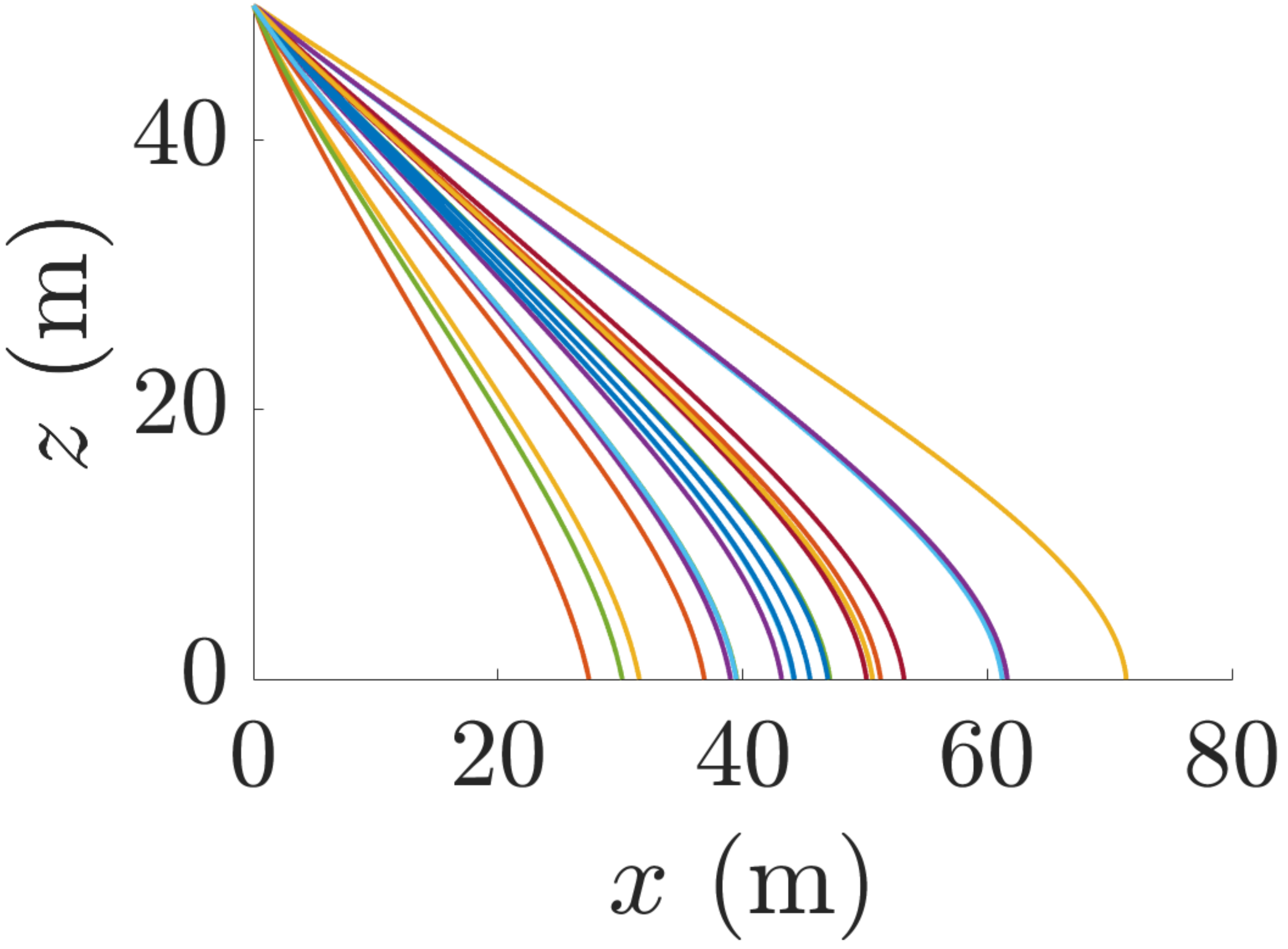}}
\subfigure[]{\label{fig:f}\includegraphics[width=0.32\textwidth]{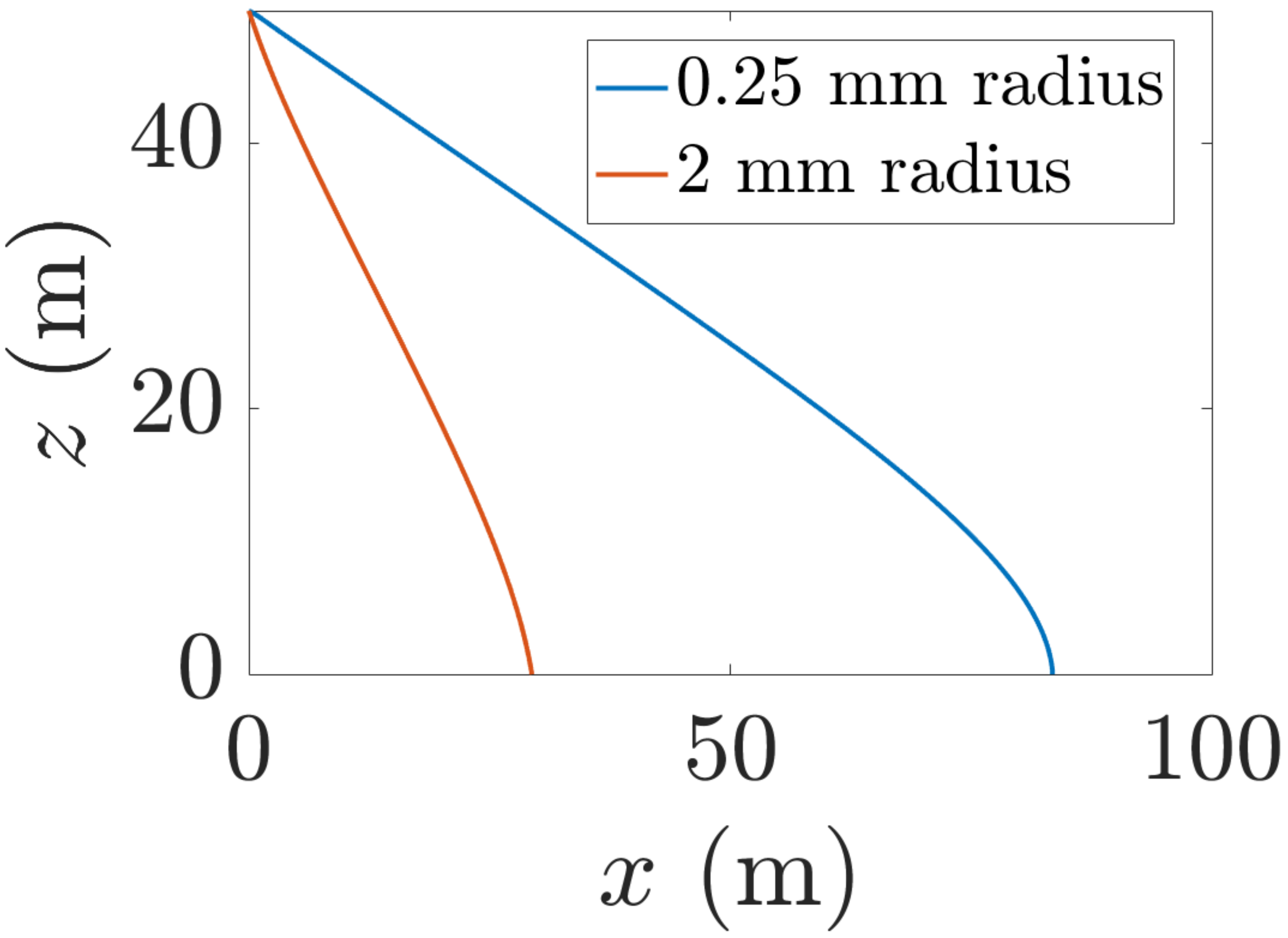}}	
\caption{Wind profile and the corresponding firebrand trajectories. 
	(a,b,c) The logarithmic velocity field~\eqref{eq:logwind}.
	(d,e,f) The hyperbolic tangent velocity field~\eqref{eq:tanhwind}.
}
\label{fig:profiles}
\end{figure}

For each method, we present our results corresponding to the logarithmic
wind profile~\eqref{eq:logwind} and the hyperbolic wind profile~\eqref{eq:tanhwind}.
Figure~\ref{fig:profiles} shows the wind profiles along with a set of corresponding firebrand trajectories. Smaller firebrands travel farther before landing.

\begin{figure}
\centering
\subfigure[]{\label{fig:LDTlog}\includegraphics[width=0.49\textwidth]{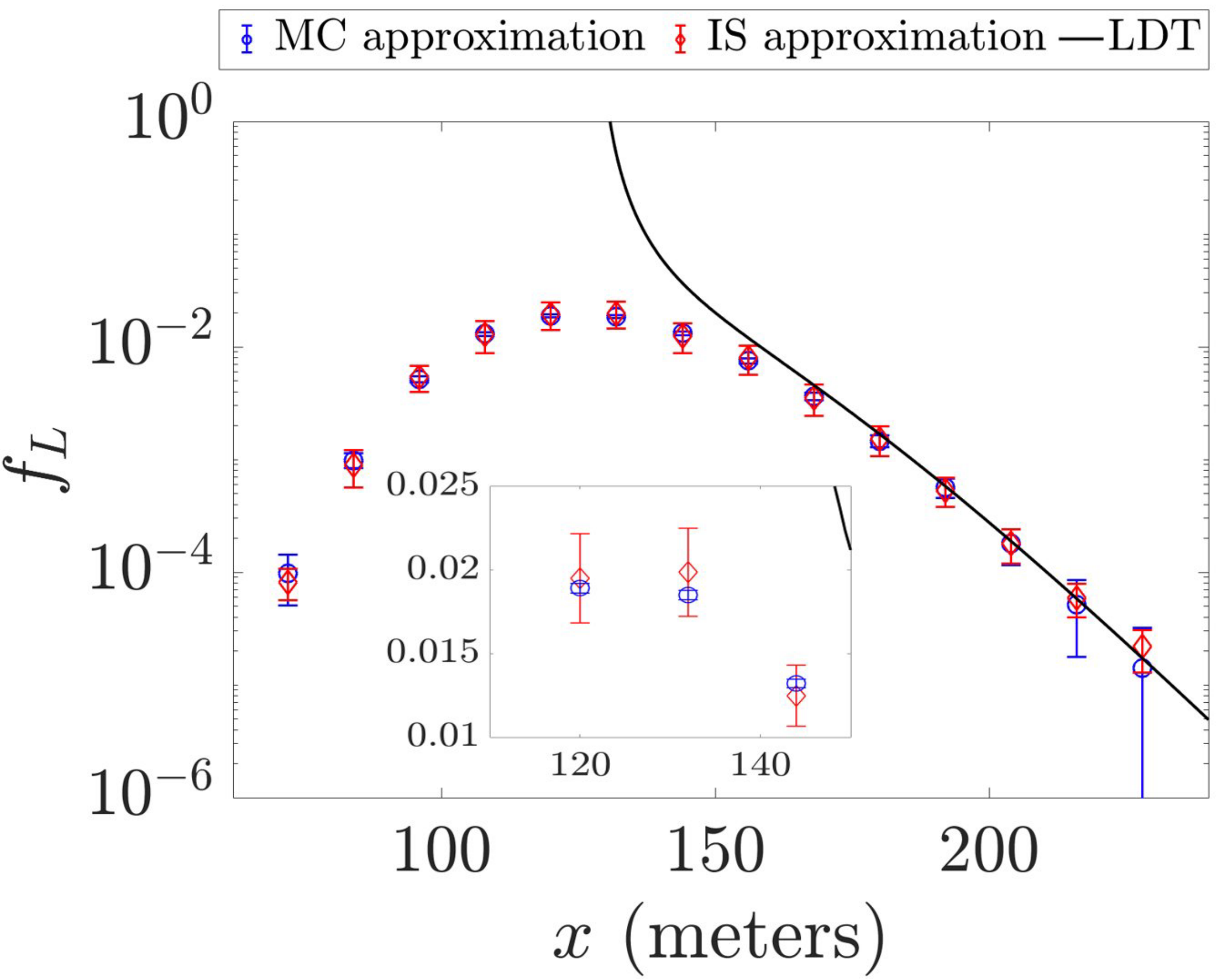}}
\subfigure[]{\label{fig:LDTtanh}\includegraphics[width=0.49\textwidth]{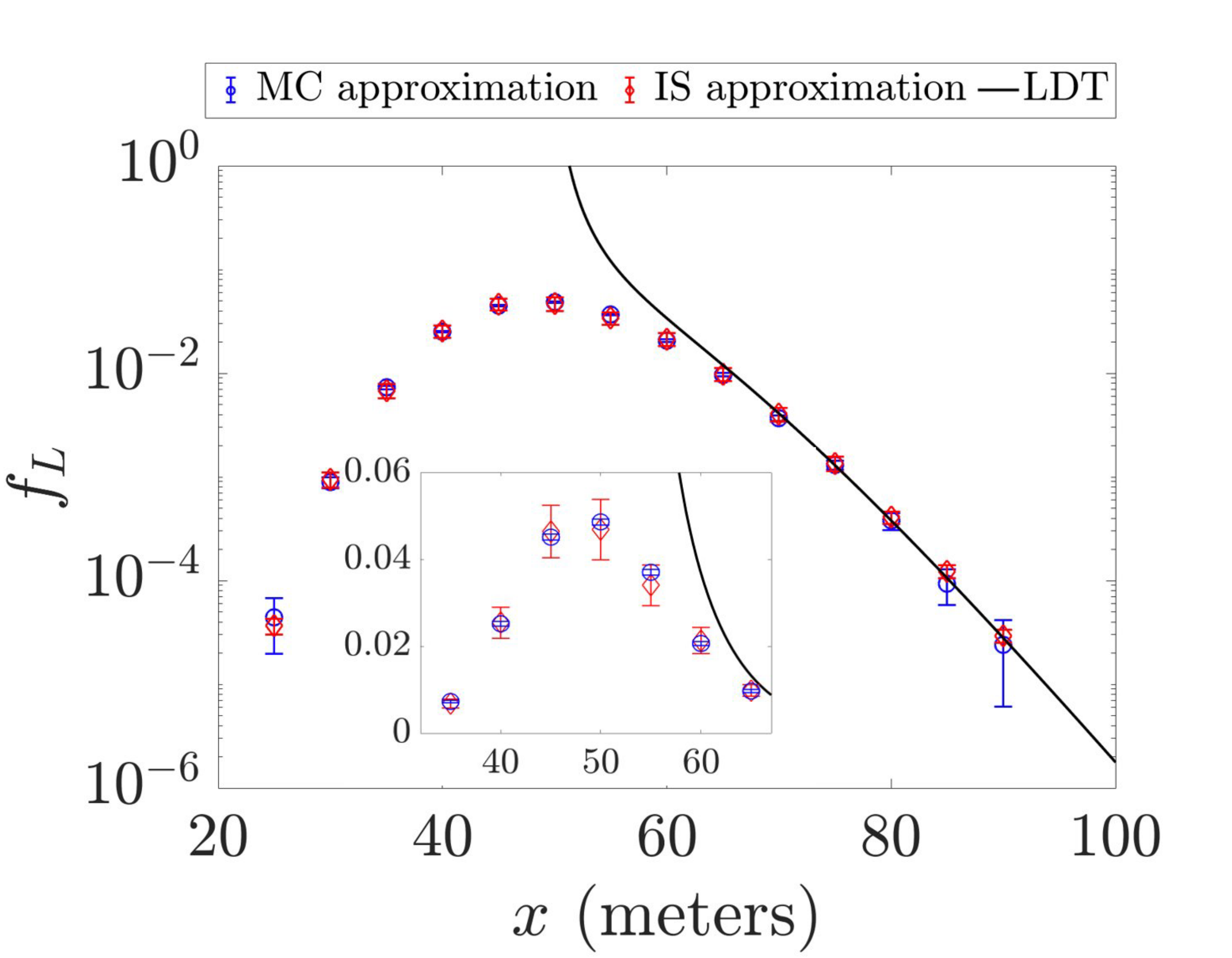}}
\caption{Landing PDF approximated with MC, IS and LDT methods. Bars represent 95 percent confidence intervals. (a) Logarithmic wind. (b) Hyperbolic tangent wind.}
\end{figure}

We begin by comparing the MC, IS, and LDT methods for quantifying the landing distance PDF $f_L$.
Figure~\ref{fig:LDTlog} shows the estimated PDF for the logarithmic wind profile.
For landing distances with relatively high probability, roughly between $100-180$ meters, crude MC provides slightly more accurate approximations as compared to IS. The inset of figure~\ref{fig:LDTlog} shows a closeup view of this high-probability region showing the smaller variance associated with the MC method.
This is due to the large number of MC realizations that land in this interval. However, when approximating low probability events at the tails of the distribution, the
IS method is more accurate as it exhibits lower variance compared to MC. Both MC and IS methods fail beyond landing distance $\ell = 230$ meters, since no samples reach beyond that landing distance. Since LDT is not based on sampling, it avoids this issue and gives approximations past the point where MC and IS fail. However, as noted in section~\ref{sec:ldt}, LDT is an asymptotic estimate and therefore it is only valid for large enough landing distance $\ell$. As a result, it fails near the mode (i.e., high probability region) of the PDF but accurately estimates the tail for $\ell>170$ meters.

Similar observations are made for landing PDF $f_L$ corresponding to the hyperbolic tangent wind profile as shown in figure~\ref{fig:LDTtanh}. Again, we see that MC produces an approximation with a smaller variance than IS at landing locations with high probability, but IS approximations have smaller variance towards the tail end of the landing distribution. For $\ell>90$ meters, there are no samples and therefore both MC and IS methods fail to produce an approximation for the PDF. In contrast, the LDT approximation can easily be extended beyond $\ell =90$ meters. The asymptotic range over which LDT is valid is over $\ell>  70$ meters.

In order to compare the computational cost of the methods, we count the number of firebrand advections and record the total computational time. The number of advections accounts for the number of samples as well as the advections needed for the optimization steps inside IS and LDT algorithms.
All methods were implemented in MATLAB, version 2018a, and executed on 12 cores of Intel Xeon CPU E5-2690 8-core CPUs with 2.90GHz DDR3 RAM. 
Firebrand trajectories are found by numerically integrating the equations of motion~\eqref{eq:motion} using the Runge--Kutta scheme of ODE45. The optimization problems in the IS and LDT methods are solved using MATLAB's built-in package fmincon.

MC approximation requires solving equation~\eqref{eq:motion} for $N_{MC}$ realizations of the firebrand size, which are then used in approximation~\eqref{eq:MCapprox}. For IS, the first step is to integrate $N_{IS}$ realizations of the firebrand size in order to set up the optimization problem~\eqref{eq:varopt}. Recall that $N_{IS}$ is much smaller than the MC sample size  $N_{MC}$. Then we choose $K$ landing distances $\ell_i$, $i=1,2,\cdots,K$. For each landing interval $D(\ell_i)$, we solve the optimization problem~\eqref{eq:varopt} to obtain a proposal distribution $q(r;\theta_\ast)$. The average number of optimization iterations are denoted by $O_{IS}$. Then $\hat N_{IS}$ realizations are taken from each proposal distribution $q$, which are then used to compute the IS approximation~\eqref{eq:ISapprox}.
For LDT, we need $N_\lambda$ values of the Lagrange multiplier $\lambda$. For each value of $\lambda$, we solve the optimization problem~\eqref{eq:unconstrained} which takes an average of $O_{LD}$ iterations to converge. This yields  $N_\lambda\times O_{LD}$ advections on average.

Table~\ref{tab:compute} compares the computational cost of MC, IS and LDT for the hyperbolic tangent wind field~\eqref{eq:tanhwind}. The results are similar for the logarithmic wind. For the MC method, we use $n_{MC} = 10^6$ samples width $K=15$ spatial intervals to estimate the landing PDF. In this sample, firebrands do not travel much farther than $\ell\simeq 90$ meters. For IS method, we use $N_{IS}=2\times 10^5$ samples from the nominal distribution $p$. 
As in MC, we use $K=15$ spatial intervals. The corresponding optimization problem takes an average of $O_{IS} = 35$ iterations to converge. Once the proposal distribution $q$ is found, we take $\hat N_{IS} = 10^3$ samples for each spatial interval. For the LDT approximation, the number of $\lambda$ values used in the sequence is $N_\lambda = 296$ and it takes an average of $O_{IS}= 45$ iterations to solve the optimization problem~\eqref{eq:unconstrained} for each value of $\lambda$.

The last column of Table~\ref{tab:compute} shows the total computational time for each method. The MC simulations are most expensive and take about 18 minutes. We note that this computational time will increase significantly in the realistic situation where the wind field is not available analytically, and a CFD simulation is needed to obtain it.
The computational time of IS is an order of magnitude smaller around 3.7 minutes. Note that the main computational cost of IS is associated with advecting the sample from the nominal distribution $p$ which takes about $204$ seconds.
LDT takes only 23 seconds and therefore is computationally most efficient. However, we reiterate that LDT is only valid for rare events at the tail of the distribution and fails to quantify the most probable events.

\begin{table}[!htb]
\centering
\begin{tabular}{|llll|}
	\hline
	Method & \begin{tabular} {@{}c@{}} Formula for \\ number of advections \end{tabular} & \begin{tabular} {@{}c@{}} Number of \\ advections \end{tabular} & Compute time  \\ \hline
	MC & $N_{MC}$ & $1,000,000$ & $1084$ sec  \\
	IS & $N_{IS} + K(\hat N_{IS}+O_{IS})$ & $215,525$ & $204 + 19$ sec \\
	LDT & $N_{\lambda}\times O_{LD}$ & $13,294$ & $23$ sec \\ \hline
\end{tabular}
\caption{Computational cost of the methods for the hyperbolic tangent wind field. The number of samples are denoted by $N_{MC}$, $N_{IS}$ and $\hat N_{IS}$. The number of intervals is denoted by $K$, and the number of optimization iterations by $O_{IS}$ and $O_{LD}$. For LDT, the number of Lagrange multipliers is given by $N_\lambda$.}
\label{tab:compute}
\end{table}

\begin{figure}
\centering
\subfigure[]{\label{fig:logmasspdf}\includegraphics[width=0.49\textwidth]{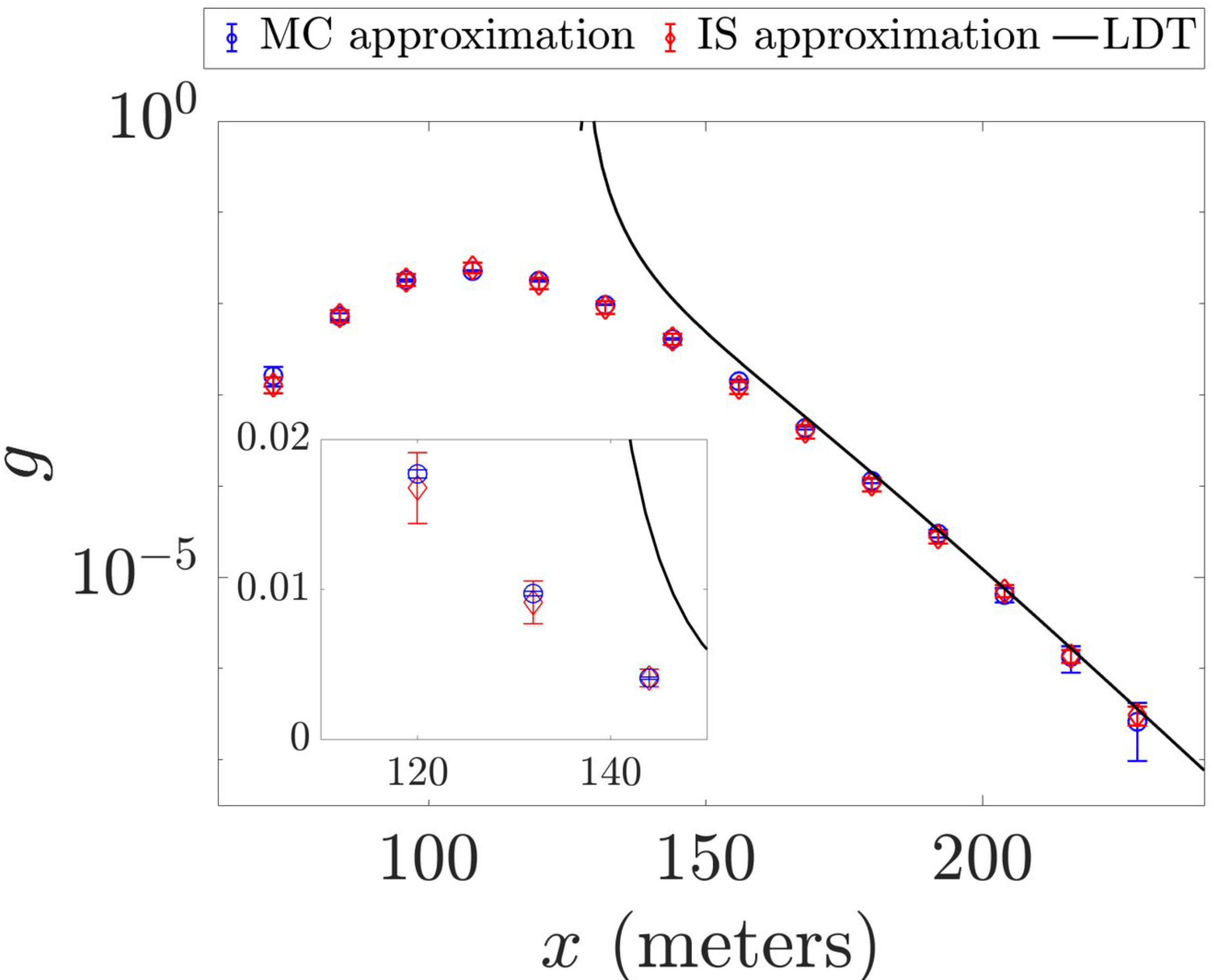}}
\subfigure[]{\label{fig:tanhmasspdf}\includegraphics[width=0.49\textwidth]{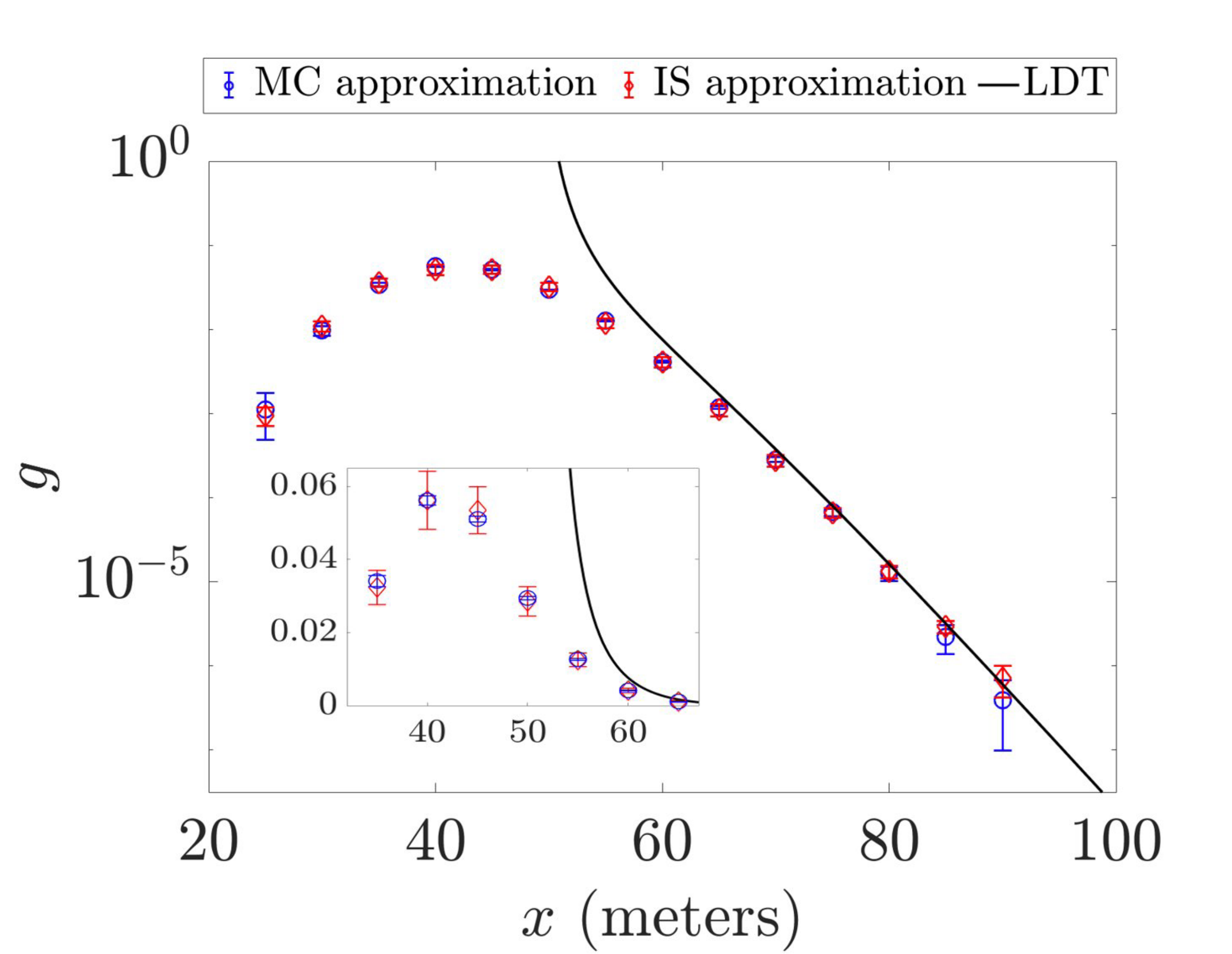}}
\caption{The relative landed mass distribution approximated by MC, IS and LDT methods. Bars represent 95 percent confidence intervals. (a) Logarithmic wind. (b) Hyperbolic tangent wind.}
\label{fig:masspdf}
\end{figure}

As mentioned in section~\ref{sec:lmd}, the relative landed mass distribution $g$ can also be computed from the landing distribution $f_L$; see equation~\eqref{eq:lmd_equiv}. Figure~\ref{fig:masspdf} shows the results for both wind fields. As expected, the relative landed mass distribution $g$ is slightly different from the landing distribution $f_L$. In particular, the tail of the relative landed mass distribution $g$ decays more rapidly and its mode occurs at a slightly smaller distance as compared to the landing distribution $f_L$. Both these features are associated with the fact that firebrands which travel a longer distance are airborne for a longer period of time and therefore burn more mass before landing.
In terms of the approximation methods, we reach the same conclusions as when approximating the landing distribution. Namely, near the mode of the distribution, MC is more accurate than IS owing to the large number of firebrands that land there. However, near the tail where low probability events occur, IS becomes more accurate than MC. Near the tail, LDT is in excellent agreement with IS results, but it fails near the mode of the distribution. 

\begin{figure}
\centering
\subfigure[]{\label{fig:vstarmostprob}\includegraphics[width=0.49\textwidth]{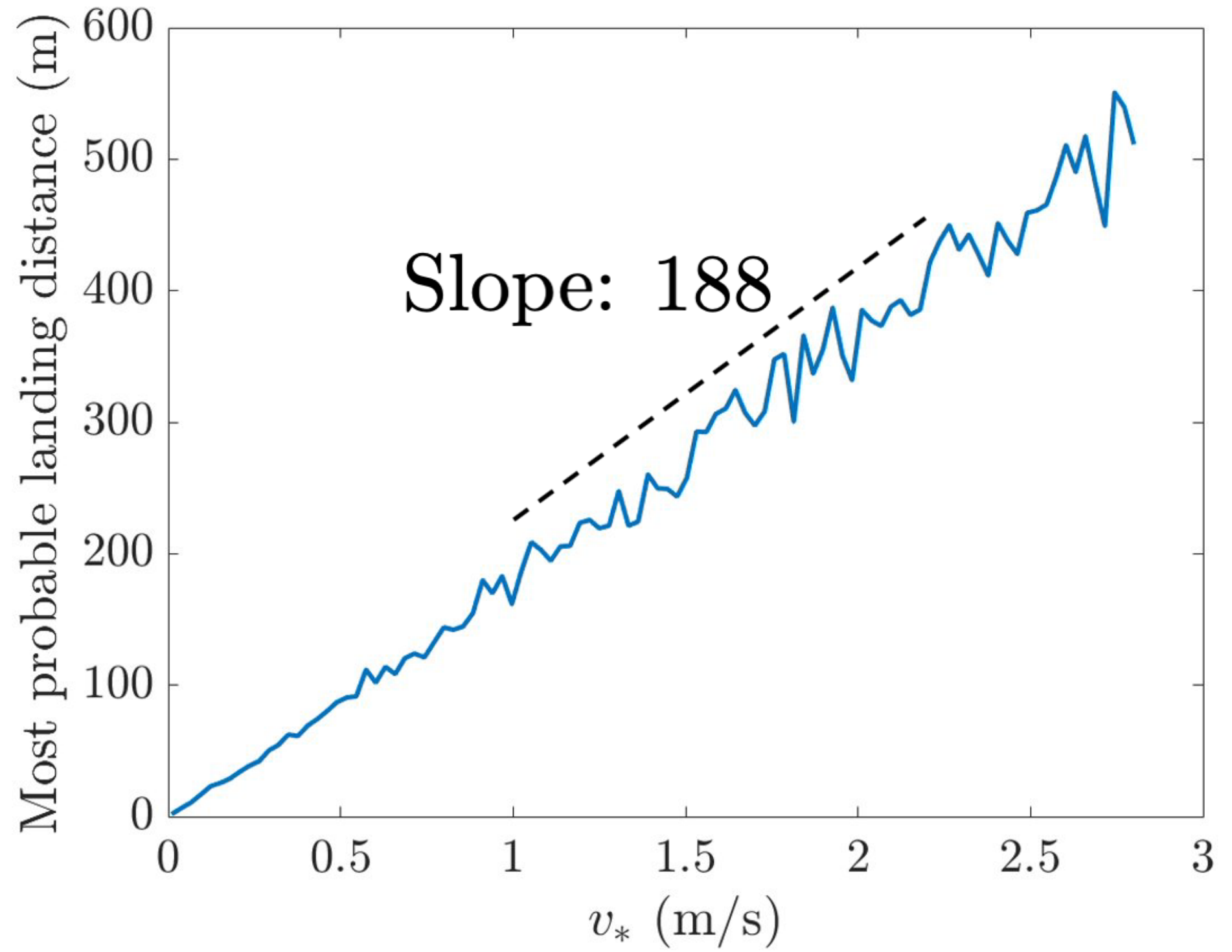}}
\subfigure[]{\label{fig:Umostprob}\includegraphics[width=0.49\textwidth]{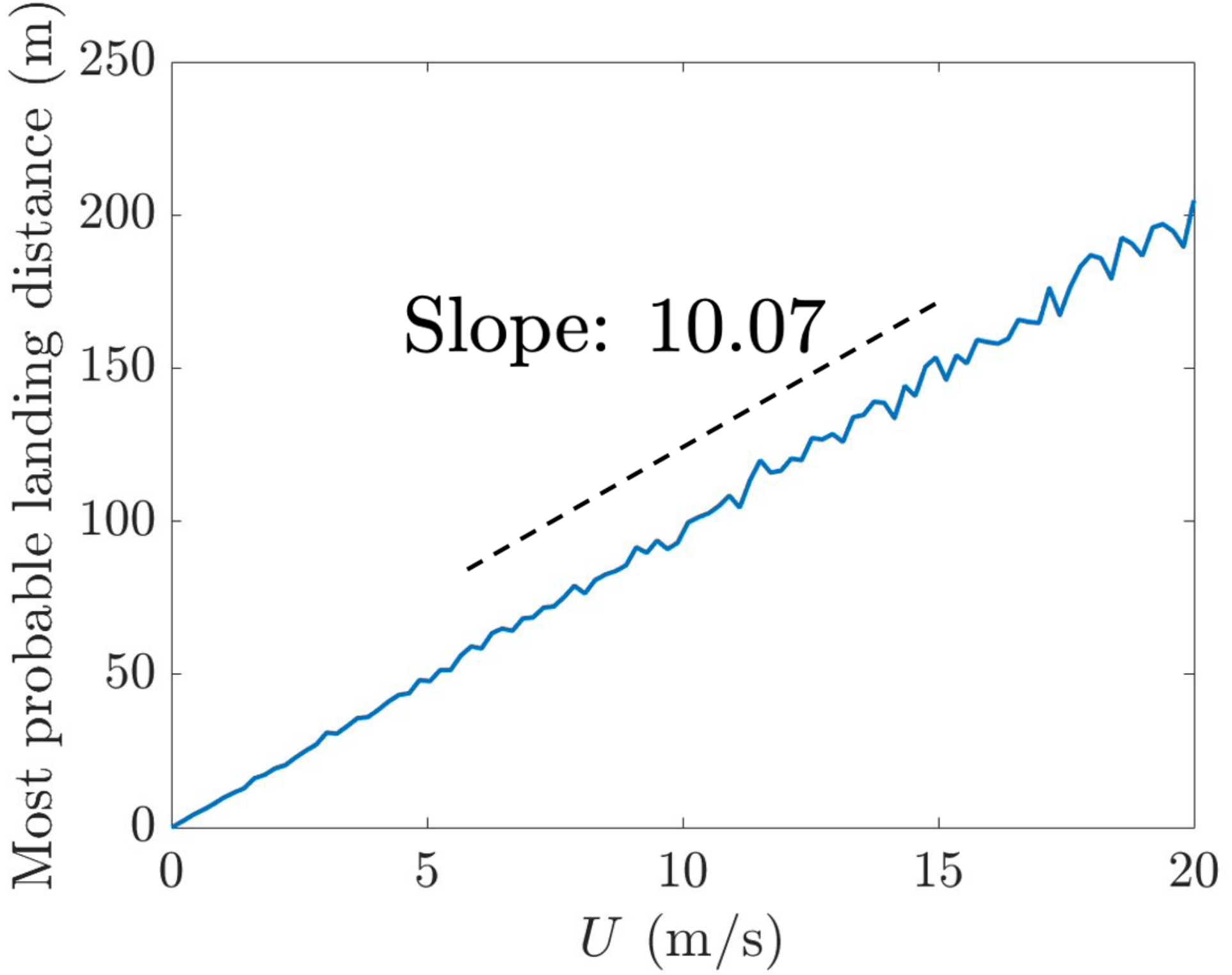}}
\caption{Most probable landing distance as a function of the characteristic velocity. 
	(a) Logarithmic wind where $v_*$ denotes the friction velocity.
	(b) Hyperbolic tangent wind where $U$ denotes the free stream velocity.}
\label{fig:peakprob}
\end{figure}

Finally, we examine the most likely landing location as a function of the characteristic velocity of the wind fields~\eqref{eq:logwind} and~\eqref{eq:tanhwind}.
So far, we have mainly focused on the tail of the distributions since they correspond to rare but consequential spotting events. Although quantifying these rare events is important, the most likely place where a spot fire can form is near the mode of the 
landing or relative landed mass distributions.
Figure~\ref{fig:peakprob} shows the most probable landing location $L_{m}$, i.e. mode of the distribution, as a function of the characteristic velocity of the logarithmic and hyperbolic tangent winds. For the logarithmic wind, $L_m$ grows linearly with the friction velocity $v_\ast$ such that $L_m\simeq 188v_\ast$. Similarly, for the hyperbolic tangent wind, we see a linear relationship
$L_m\simeq 10U$.
In both cases, the mode grows linearly with the characteristic velocity which is notable because the equations of motion~\eqref{eq:motion} depend quadratically  on the relative velocity $\vc u-\vc v$.

\section{Conclusions}
\label{sec:conclusion}
The main objective of this study was to quantify the fire spotting events by approximating the landing distribution and relative landed mass distribution of firebrands. This is a challenging problem because the resulting distributions have heavy tails, corresponding to rare but consequential spotting events far away from the main fire. We compared three different methods: crude Monte Carlo simulations, importance sampling, and large deviation theory. 

The MC method is most expensive and returns the least accurate tail approximations. Since large landing distances are rare, a very large sample size is required to approximate the tail with a reasonable accuracy. Here, for simplicity, we used analytically prescribed wind fields. But in realistic situations, where the wind field is obtained from a CFD model, such large sample sizes are not practical.

We then turned to the variance reduction method of importance sampling. IS uses a small initial MC sample to obtain the nominal distribution. Then the IS proposal distributions are computed by solving an optimization problem. The total number of realizations used for IS was an order of magnitude smaller than MC, yet IS approximations were more accurate in quantifying the tail of the distribution. More precisely, as the landing distance grows, the variance of the MC approximation also grows. However, the variance of IS remains bounded and relatively small. We note that, although IS is more accurate at the tails, the MC method is slightly more accurate near the mode of the distribution owing to the large number of firebrands that land there.

The third method we considered is a based on recently formulated large deviation theory for ordinary and stochastic differential equations~\cite{Dematteis2019}. LDT offers an asymptotic approximation for the distribution in terms of the so-called rate function. To evaluate the rate function, one needs to solve an associated optimization problem. Unlike MC and IS, LDT does not require any sampling. As a result, LDT is computationally  more efficient than both MC and IS, being two and one order of magnitude faster, respectively. Furthermore, MC and IS are both limited by the farthest landing firebrand from a sample such that the tail distribution cannot be approximated beyond this point. But since LDT does not rely on sampling, it can approximate the tail at arbitrarily large distances.

The LDT method has two notable drawbacks. First, being an asymptotic theory, it can only quantify the tail of the distribution and fails to approximate the mode. Second, to the best of our knowledge, the current large deviation theory is not equipped with error bars to quantify the accuracy of the approximation. Although our results show that LDT results are in excellent agreement with MC and IS results, the accuracy cannot be known a priori.

Given our observations, we recommend a hybrid approach for quantifying spotting distributions where the MC or IS methods are used to quantify the high-probability events near the mode of the distribution. Accurate results can be obtained here even with a relatively small sample size since most firebrands land near the mode. In contrast, LDT method should be used to quantify low probability events at the tails. Since LDT does not require any sampling, it quantifies the tails accurately at a fraction of the computational cost.

We also considered the relative landed mass distribution, which quantifies the proportion of the firebrand mass landed at a distance. 
Since firebrands burn throughout their flight, the relative landed mass is generally different from the landing distribution. Nonetheless, in section~\ref{sec:lmd}, we derived a formula which allowed us to compute the relative landed mass distribution from the landing distribution, at no significant computational cost.

In addition to approximating probability distributions, we also examined the effect of the wind field on the landing distance. Specifically, the relationship between the most probable landing distance and the characteristic velocity of the wind was observed to be linear. This is despite the nonlinearities in the equations of motion.

Future work will focus on relaxing the simplifying assumptions made in this paper (see section \ref{sec:assump}), with the ultimate goal of implementing an efficient spotting quantifier in high-fidelity fire simulators such as HIGRAD/FIRETEC, QUIC-FIRE, and WRF-SFIRE. Existing work on inertial particle transport~\cite{koo2010, haller2008, beron2015} suggests that, in turbulent flows, firebrands get trapped inside vortices and therefore quantifying their landing distribution may present new challenges that are absent in laminar steady flows.

\section*{Declaration of competing interest}
The authors declare that they have no known competing financial
interests or personal relationships that could have appeared to influence
the work reported in this paper.

\section*{Acknowledgments}
We are grateful to Dr. Joseph O'Brien (USDA Forest Service) for fruitful conversations.

\section*{Funding}
This work was partially supported by the National Science Foundation grant DMS-1745654.


\begin{thebibliography}{10}
	
	\bibitem{Tedim2018}
	F.~Tedim, V.~Leone, M.~Amraoui, C.~Bouillon, M.~R. Coughlan, G.~M. Delogu,
	P.~M. Fernandes, C.~Ferreira, S.~McCaffrey, T.~K. McGee, J.~Parente,
	D.~Paton, M.~G. Pereira, L.~M. Ribeiro, D.~X. Viegas, and G.~Xanthopoulos.
	\newblock Defining extreme wildfire events: {D}ifficulties, challenges, and
	impacts.
	\newblock {\em Fire}, 1(1), 2018.
	
	\bibitem{tohidi2017}
	A.~Tohidi and N.B. Kaye.
	\newblock Aerodynamic characterization of rod-like debris with application to
	firebrand transport.
	\newblock {\em Journal of Wind Engineering and Industrial Aerodynamics},
	168:297--311, 2017.
	
	\bibitem{Mendez2021}
	A.~Mendez and M.~Farazmand.
	\newblock Investigating climate tipping points under various emission reduction
	and carbon capture scenarios with a stochastic climate model.
	\newblock {\em Proceedings of the Royal Society A: Mathematical, Physical and
		Engineering Sciences}, 477(2256):20210697, 2021.
	
	\bibitem{burke2021}
	M.~Burke, A.~Driscoll, S.~Heft-Neal, J.~Xue, J.~Burney, and M.~Wara.
	\newblock The changing risk and burden of wildfire in the {U}nited {S}tates.
	\newblock {\em Proceedings of the National Academy of Sciences},
	118(2):e2011048118, 2021.
	
	\bibitem{albini1979}
	F.~A. Albini.
	\newblock Spot fire distance from burning trees - {A} predictive model.
	\newblock Technical report, US Forest Service, 1979.
	
	\bibitem{koo2010}
	E.~Koo, P.~J. Pagni, D.~R. Weise, and J.~P. Woycheese.
	\newblock Firebrands and spotting ignition in large-scale fires.
	\newblock {\em International Journal of Wildland Fire}, 19(7):818--843, 2010.
	
	\bibitem{tarifa1965}
	C.~S. Tarifa, P.~P del Notario, A.~R. Villa, M.~L. Martinez, and O.~Perez.
	\newblock Open fires and transport of firebrands.
	\newblock Technical report, Instituto Nacional De Tecnica Aeroespacial, 1965.
	
	\bibitem{weir2007}
	J.~R. Weir.
	\newblock Using relative humidity to predict spotfire probability on prescribed
	burns.
	\newblock {\em In: Sosebee, Ronald E.; Wester, David B.; Britton, Carlton M.;
		McArthur, E. Durant; Kitchen, Stanley G., comps. Proceedings: Shrubland
		dynamics--fire and water; 2004 August 10-12; Lubbock, TX. Proceedings
		RMRS-P-47. Fort Collins, CO: US Department of Agriculture, Forest Service,
		Rocky Mountain Research Station. p. 69-72.}, 47, 2007.
	
	\bibitem{bucklew2004}
	J.~A. Bucklew.
	\newblock {\em Introduction to Rare Event Simulation}.
	\newblock Springer Series in Statistics. Springer, New York, USA, 2004.
	
	\bibitem{farazmand2019a}
	M.~Farazmand and T.~P. Sapsis.
	\newblock {Extreme {E}vents: {M}echanisms and {P}rediction}.
	\newblock {\em Applied Mechanics Reviews}, 71(5), 2019.
	
	\bibitem{Dematteis2018}
	G.~Dematteis, T.~Grafke, and E.~Vanden-Eijnden.
	\newblock Rogue waves and large deviations in deep sea.
	\newblock {\em Proceedings of the National Academy of Sciences},
	115(5):855--860, 2018.
	
	\bibitem{Dematteis2019}
	G.~Dematteis, T.~Grafke, and E.~Vanden-Eijnden.
	\newblock Extreme event quantification in dynamical systems with random
	components.
	\newblock {\em SIAM/ASA Journal on Uncertainty Quantification},
	7(3):1029--1059, 2019.
	
	\bibitem{tong2021}
	S.~Tong, E.~Vanden-{E}ijnden, and G.~Stadler.
	\newblock Extreme event probability estimation using {PDE}-constrained
	optimization and large deviation theory, with application to tsunamis.
	\newblock {\em Communications in Applied Mathematics and Computational
		Science}, 16(2):181--225, 2021.
	
	\bibitem{sardoy2007}
	N.~Sardoy, J-L. Consalvi, B.~Porterie, and A.~Fernandez-Pello.
	\newblock Modeling transport and combustion of firebrands from burning trees.
	\newblock {\em Combustion and Flame}, 150:151--169, 08 2007.
	
	\bibitem{bhutia2010}
	S.~Bhutia, M.~Jenkins, and R.~Sun.
	\newblock Comparison of firebrand propagation prediction by a plume model and
	coupled-fire/atmosphere large-eddy simulator.
	\newblock {\em Journal of Advances in Modeling Earth Systems}, 2(1):4, 2010.
	
	\bibitem{jenkins2009}
	R.~Sun, S.~Krueger, M.~Jenkins, M.~Zulauf, and J.~Charney.
	\newblock The importance of fire-atmosphere coupling and boundary-layer
	turbulence to wildfire spread.
	\newblock {\em International Journal of Wildland Fire}, 18(50-60), 01 2009.
	
	\bibitem{anand2018}
	C.~Anand, B.~Shotorban, and S.~Mahalingam.
	\newblock Dispersion and deposition of firebrands in a turbulent boundary
	layer.
	\newblock {\em International Journal of Multiphase Flow}, 109:98--113, 2018.
	
	\bibitem{manzello2020}
	S.~L. Manzello, S.~Suzuki, M.J. Gollner, and A.~C. Fernandez-Pello.
	\newblock Role of firebrand combustion in large outdoor fire spread.
	\newblock {\em Progress in Energy and Combustion Science}, 76:100801, 2020.
	
	\bibitem{Koo2012}
	E.~Koo, R.~R. Linn, P.~J. Pagni, and C.~B. Edminster.
	\newblock {{Modelling firebrand transport in wildfires using HIGRAD/FIRETEC}}.
	\newblock {\em International Journal of Wildland Fire}, 21:396--417, 2012.
	
	\bibitem{quicfire}
	R.~R. Linn, S.~L. Goodrick, S.~Brambilla, M.~J. Brown, R.~S. Middleton, J.~J.
	O'Brien, and J.~K. Hiers.
	\newblock {{QUIC-fire: A fast-running simulation tool for prescribed fire
			planning}}.
	\newblock {\em Environmental Modelling and Software}, 125:104616, 2020.
	
	\bibitem{wrfsfire}
	J.~Mandel, J.~D. Beezley, and A.~K. Kochanski.
	\newblock {{Coupled atmosphere-wildland fire modeling with WRF 3.3 and SFIRE
			2011}}.
	\newblock {\em Geosci. Model Dev.}, 4:591--610, 2011.
	
	\bibitem{Tarifa1963}
	C.~S. Tarifa, P.~P. del Notario, and F.~G. Moreno.
	\newblock On flight paths and lifetimes of burning particles of wood.
	\newblock {\em Symposium (International) on Combustion}, 10(1):1021--1037,
	1965.
	
	\bibitem{Lee1969}
	S.~L. Lee and J.~M. Hellman.
	\newblock Study of firebrand trajectories in a turbulent swirling natural
	convection plume.
	\newblock {\em Combustion and Flame}, 13(6):645--655, 1969.
	
	\bibitem{tse1998}
	S.~D. Tse and A.~C. Fernandez-Pello.
	\newblock {{On the flight paths of metal particles and embers generated by
			power lines in high-winds - A potential source of wildland fires}}.
	\newblock {\em Fire Safety Journal}, 30:333--356, 1998.
	
	\bibitem{zheng2020}
	G.~Zheng, A.~J. Sedlacek, A.~C. Aiken, Y.~Feng, T.~B. Watson, S.~Raveh-Rubin,
	J.~Uin, E.~R. Lewis, and J.~Wang.
	\newblock Long-range transported {N}orth {A}merican wildfire aerosols observed
	in marine boundary layer of eastern {N}orth {A}tlantic.
	\newblock {\em Environmental International}, 139:105680, 2020.
	
	\bibitem{Albini2012}
	Albini~F. A., Alexander~M. E., and Cruz~M. G.
	\newblock A mathematical model for predicting the maximum potential spotting
	distance from a crown fire.
	\newblock {\em International Journal of Wildland Fire}, 21:609--627, 2012.
	
	\bibitem{clarke1994}
	K.~C. Clarke, J.~A. Brass, and P.~J. Riggan.
	\newblock A cellular automaton model of wildfire propagation and extinction.
	\newblock {\em Photogrammetric Engineering and Remote Sensing},
	60(11):1355--1367, 1994.
	
	\bibitem{Karafyllidis1997}
	I.~Karafyllidis and A.~Thanailakis.
	\newblock A model for predicting forest fire spreading using cellular automata.
	\newblock {\em Ecological Modelling}, 99(1):87--97, 1997.
	
	\bibitem{boychuk2007}
	D.~Boychuk, W.~J. Braun, R.~J. Kulperger, Z.~L. Krougly, and D.~A. Stanford.
	\newblock A stochastic model for forest fire growth.
	\newblock {\em INFOR: Information Systems and Operational Research},
	45(1):9--16, 2007.
	
	\bibitem{Alexandridis2008}
	A.~Alexandridis, D.~Vakalis, C.I. Siettos, and G.V. Bafas.
	\newblock A cellular automata model for forest fire spread prediction: {T}he
	case of the wildfire that swept through {S}petses {I}sland in 1990.
	\newblock {\em Applied Mathematics and Computation}, 204(1):191--201, 2008.
	
	\bibitem{Hillen2015}
	T.~Hillen, B.~Greese, J.~Martin, and G.~de~Vries.
	\newblock Birth-jump processes and application to forest fire spotting.
	\newblock {\em Journal of Biological Dynamics}, 9(sup1):104--127, 2015.
	
	\bibitem{Hillen2016}
	J.~Martin and T.~Hillen.
	\newblock The spotting distribution of wildfires.
	\newblock {\em Applied Sciences}, 6(6):177, 2016.
	
	\bibitem{maxey1983}
	M.~R. Maxey and J.J. Riley.
	\newblock Equation of motion for a small rigid sphere in a nonuniform flow.
	\newblock {\em Phys. Fluids}, 26:883--889, 1983.
	
	\bibitem{kim1998}
	I.~Kim, S.~Elghobashi, and W.~A. Sirignano.
	\newblock On the equation for spherical-particle motion: effect of reynolds and
	acceleration numbers.
	\newblock {\em Journal of Fluid Mechanics}, 367:221--253, 1998.
	
	\bibitem{babiano2000}
	A.~Babiano, J.~H.~E. Cartwright, O.~Piro, and A.~Provenzale.
	\newblock {{Dynamics of a small neutrally buoyant sphere in a fluid and
			targeting in Hamiltonian systems}}.
	\newblock {\em Phys. Rev. Lett.}, 84:5764--5767, 2000.
	
	\bibitem{linn2007}
	R.~R. Linn, E.~Koo, C.~B. Edminster, and J.~A. Perry.
	\newblock Using a physics-based model to characterize spotting potential for
	protection of the wildland urban interface.
	\newblock Technical Report 07-1-5-01, Joint Fire Science Program, 2007.
	
	\bibitem{MR_EUR}
	M.~Farazmand and G.~Haller.
	\newblock The {M}axey--{R}iley equation: {E}xistence, uniqueness and regularity
	of solutions.
	\newblock {\em Nonlinear Analysis: Real World Applications}, 22:98--106, 2015.
	
	\bibitem{langlois2015}
	G.~P. Langlois, M.~Farazmand, and G.~Haller.
	\newblock Asymptotic dynamics of inertial particles with memory.
	\newblock {\em Journal of nonlinear science}, 25(6):1225--1255, 2015.
	
	\bibitem{sommerfeld1952}
	A.~Sommerfeld.
	\newblock {\em Mechanics: {L}ectures on theoretical physics}, volume~1.
	\newblock Academic press, 1952.
	
	\bibitem{plastino1992}
	A.~R. Plastino and J.~C. Muzzio.
	\newblock On the use and abuse of {N}ewton's second law for variable mass
	problems.
	\newblock {\em Celestial Mechanics and Dynamical Astronomy}, 53(3):227--232,
	1992.
	
	\bibitem{martin2013}
	J.~Martin.
	\newblock {\em Derivation and investigation of mathematical models for spotting
		in wildland fire}.
	\newblock PhD thesis, University of Alberta, 2013.
	
	\bibitem{bee2009}
	M.~Bee.
	\newblock Importance sampling for sums of lognormal distributions with
	applications to operational risk.
	\newblock {\em Communications in Statistics - Simulation and Computation},
	38(5):939--960, 2009.
	
	\bibitem{kloek1978}
	T.~Kloek and H.~K. van Dijk.
	\newblock {{Bayesian Estimates of Equation System Parameters: An Application of
			Integration by Monte Carlo}}.
	\newblock {\em Econometrica}, 46(1):1--19, Jan 1978.
	
	\bibitem{mcleish2015}
	D.~L. McLeish and Z.~Men.
	\newblock Extreme value importance sampling for rare event risk measurement.
	\newblock {\em Innovations in Quantitative Risk Management}, 99, 2015.
	
	\bibitem{varadhan1984}
	S.~R.~S. Varadhan.
	\newblock {\em Large deviations and applications}.
	\newblock SIAM, Philadelphia, Pennsylvania, 1984.
	
	\bibitem{hollander2000}
	F.~den Hollander.
	\newblock {\em Large Deviations}.
	\newblock Fields Institute Monographs. American Mathematical Society,
	Providence, Rhod e Island, 2000.
	
	\bibitem{dembo1998}
	A.~Dembo and O.~Zeitouni.
	\newblock {\em Large Deviations Techniques and Applications}.
	\newblock Stochastic Modelling and Applied Probability. Springer, Berlin,
	Germany, second edition, 1998.
	
	\bibitem{haller2008}
	G.~Haller and T.~Sapsis.
	\newblock Where do inertial particles go in fluid flows?
	\newblock {\em Physica D}, 237(5):573--583, 2008.
	
	\bibitem{beron2015}
	F.~J. Beron-Vera, M.~J. Olascoaga, G.~Haller, M.~Farazmand, J.~Tri\~nanes, and
	Y.~Wang.
	\newblock Dissipative inertial transport patterns near coherent lagrangian
	eddies in the ocean.
	\newblock {\em Chaos: An Interdisciplinary Journal of Nonlinear Science},
	25(8):087412, 2015.
	
\end{thebibliography}

\end{document}